\documentclass[%
 aps,
 pra,%
 amsmath,amssymb,
reprint,%
]{revtex4-1}

\usepackage{graphicx}
\usepackage{dcolumn}
\usepackage{bm}

\begin{document}

\preprint{AIP/123-QED}

\title[]{A unified approach to mode splitting and scattering loss in high-$Q$ whispering-gallery-mode microresonators}

\author{Qing Li}
\author{Ali A. Eftekhar}
\author{Zhixuan Xia}
\author{Ali Adibi}
\email{adibi@ee.gatech.edu}
\affiliation{School of Electrical and Computer Engineering, Georgia Institute of Technology}

\date{\today}

\begin{abstract}
Current theoretical treatment of mode splitting and scattering loss resulting from sub-wavelength scatterers attached to the surface of high-quality-factor whispering-gallery-mode microresonators is not satisfactory. Different models have been proposed for two distinct scatterer regimes, i.e., a-few- and many-scatterers. In addition, many experimental results seem difficult to understand within the existing theoretical framework. Here we develop a unified approach that applies to an arbitrary number of scatterers, which reveals the applicable conditions and the limits of the existing theoretical models. Moreover, many new understandings on mode splitting and scattering loss have been achieved, which are supported by numerical and experimental evidences. Such a unified approach is essential for the fundamental studies as well as the practical applications of mode splitting and scattering loss in high-quality-factor whispering-gallery-mode microresonators.
\end{abstract}

\pacs{Valid PACS appear here}
\maketitle

\section{Introduction}
Whispering-gallery-mode (WGM) microresonators have attracted a lot of research interest due to their high quality factors ($Q$s) and microscale mode volumes \cite{Vahala,VS}. The high $Q/V$ factor, i.e., the so-called Purcell factor \cite{Vahala}, enables strong light-matter interactions and is the essential reason behind the wide applications of WGM microresonators including low-threshold lasing \cite{laser}, low-power optical modulation \cite{modulator}, single-nanoparticle sensing \cite{sensing1,sensing2}, ultrasensitive micromechanical displacement detection \cite{opto}, as well as the fundamental studies on cavity quantum electrodynamics \cite{kimble}. Because of the structural symmetry, ideal WGM microresonators usually exhibit degeneracies in their resonant modes. For example, clockwise (CW) and counterclockwise (CCW) WGM modes are supported in microtoroid and microdisk resonators with identical mode properties (e.g., resonance frequency and linewidth). Reflected in transmission, only one single resonance is observed.

The degeneracy in the resonant modes can be lifted by destroying the structural symmetry, either done intentionally such as by introducing nanoparticles to the surface of microresonators \cite{singleprl,double}, or caused by imperfect fabrications which distort the structure \cite{borselli}. In consequence, the two degenerate modes (i.e., the CW and CCW modes) will couple to each other and a doublet appears in the transmission, a phenomenon termed as mode splitting \cite{singleol} . The structural defects can also couple the confined WGM modes to free-space radiation modes, generating scattering loss and thus linewidth broadening to the WGM modes \cite{singleprl}.

Mode splitting and scattering loss have been investigated in many different works \cite{singleprl,double,Deych,multiple,WGMstructure, borselli}. Based on their applications and the number of scatterers, previous works can be categorized into two distinct scenarios. In the first scenario, small nanoparticles are introduced to the surface of high-$Q$ microresonators, with focused applications such as strong light-matter interactions and nanoparticle sensing \cite{sensing1,sensing2,kimble}. The number of nanoparticles is usually limited to a few, and reasonably good agreements between experimental observations and developed models have been achieved. The other scenario considers sub-wavelength scatterers that are intrinsic to microresonators, such as surface roughness caused by imperfect fabrications which is of fundamental importance for a thorough understanding of high-$Q$ microresonators \cite{borselli}. In such cases, the number of scatterers is typically on the order of hundreds or even thousands, and a different approach has to be taken to study the mode splitting and scattering loss \cite{borselli}.

As we shall detail in section II, existing theoretical works developed for the above-mentioned two scenarios have only achieved partial success. Roughly speaking, they can explain the mode splitting quite well but have major problems in predicting the corresponding scattering loss even qualitatively. Though the models developed for the a-few-scatterer case have claimed possible extensions to the many-scatterer scenario \cite{double,multiple}, we will show that they have neglected some fundamental effects and such an extension will not be successful. On the hand, the approach developed for the many-scatterer case has its own shortcomings. For example, in experiments the two split modes usually exhibit different linewidths, a fact that cannot be explained by the model shown in Ref.~\cite{borselli} in a self-consistent manner. Moreover, many other experimental observations, such as the azimuthal-order variations of the mode splitting and scattering loss within the same radial mode family in an individual microresonator \cite{qingol}, are difficult to understand within that framework.

In this paper, we will develop a unified approach that applies to an arbitrary number of scatterers. Our results are closely compared to those of the existing approaches for the two distinct (i.e., a-few- and many-scatterer) scenarios , and we show conditions under which our model can be reduced to the existing models in their respective regimes. Moreover, our work has provided new understandings on the mode splitting and scattering loss in high-$Q$ WGM microresonators, which are supported by numerical studies and experimental results. For example, there is an intuitive belief that in the presence of mode splitting, the eigenmode exhibiting a lower resonance frequency of the doublet would also incur a stronger scattering loss from positive dielectric perturbations, which is true for the single-scatterer case and also consistent with existing theoretical models \cite{singleol, multiple} (see detailed discussion in section II). However, we will prove that such perception is generally invalid. Another example is that our model predicts when the two eigenmodes overlap in the resonance frequency (i.e., no mode splitting), generally their scattering loss rates are different. Therefore, the two eigenmodes are not degenerate in the strict sense, a fact that has been overlooked in previous studies \cite{double, multiple}. Furthermore, we apply our model to the fabrication-induced surface roughness in high-$Q$ WGM microresonators, which unveils that the mode splitting for different azimuthal orders are statistically independent. Hence, for the same radial mode family in an individual microresonator, strong variations of mode splitting are possible. In addition, we have shown that the scattering loss of the same radial mode family can exhibit more than 30\% variations among different azimuthal orders, while initially one might expect a uniform scattering loss rate for these modes. This in fact solves one mystery that often confuses people working on high-$Q$ microresonators, that is, in scattering-loss-limited microresonators, the extremely high intrinsic $Q$ could only be observed for one azimuthal order while the intrinsic $Q$s of the rest modes could be much lower \cite{borselliol,qingol,mohammadoe}.

\section{Existing models versus experiments}
In this section, we will review major existing models on the mode splitting and scattering loss in high-$Q$ WGM microresonators. As mentioned, currently there are two different approaches working at two distinct regimes of scatterers. By comparing the theoretical results predicted by these models to experimental (or numerical) observations, it is shown that the two approaches are only partially successful, in the sense that they either are not self-consistent or fail to agree with some of the experimental results. Therefore, a unified approach, which could provide a full understanding on the mode splitting and scattering loss, needs to be developed.

The first approach considers a coupled system consisting of the CW and CCW WGM modes as wells as free space radiation modes, with their interactions assisted by each individual scatterer. The single-scatterer case has been well studied in high-$Q$ microtoroid resonators \cite{singleprl,Deych}. Two standing-wave modes, being symmetric and anti-symmetric combinations of the CW and CCW travelling modes, appear in the resonance spectrum. The symmetric mode has a nonzero field overlap with the scatterer, resulting in a red shift in its resonance frequency and a broadening in its linewidth (we assume positive dielectric perturbations from scatterers throughout the paper unless specified). On the other hand, the anti-symmetric mode has a zero field overlap with the scatterer; therefore, its resonance frequency and linewidth stay the same as those of the CW (CCW) mode without the scatterer. These two resonances are illustrated in Fig.~1(a), with $\omega_+$ ($\omega_-$) and $\gamma_+$ ($\gamma_-$) denote the resonance frequency and loss rate of the eigenmode that has a high (lower) resonance frequency of the two split modes (i.e., $\omega_+ \geq \omega_-$). The loss of a high-$Q$ microresonator can come from many sources, such as scattering loss due to scatterers and absorption loss from material absorption. Since we are primarily interested in scattering loss in this paper, other loss mechanisms are not considered unless specified. As illustrated in Fig.~1(a), for the single-scatterer case, we have $\gamma_+ < \gamma_-$. The two-scatterer scenario has also been explored in Ref.~\cite{double}, which has experimentally demonstrated that the two eigenmodes can either have no mode spitting ($\omega_+ = \omega_-$) or a symmetric ($\gamma_+ = \gamma_-$) or an asymmetric ($\gamma_+ < \gamma_-$) doublet, depending on the relative position of the two scatterers. These experimental results can be analyzed by a generalized model presented in Ref.~\cite{multiple}, which considers multiple scatterers that are well separated apart so that the contribution from each scatterer can be considered to be independent with each other. However, when the scatterers are closely spaced, this independent-scatterer approach fails to predict correct results. For example, for $N$ identical scatterers, the model in Ref.~\cite{multiple} gives
\begin{align}
\omega_\pm = \omega_c -&NG_0  \pm G_0\left|\sum_{n=1} ^N e^{i2kx_n} \right|,\\
\gamma_\pm =N&\Gamma_0\mp \Gamma_0\left|\sum_{n=1} ^N e^{i2kx_n} \right|,
\end{align}
where $\omega_c$ is the originally degenerate resonance frequency of the WGM modes before the introduction of scatterers;   $G_0$ and $\Gamma_0$ are parameters (both positive) characterizing the resonance frequency shift and linewidth broadening caused by an individual scatterer, respectively; $k$ is the wavenumber of the WGM mode; and $x_n$ is the projection of the $n$th scatterer's position on the WGM's wave travelling direction.
From Eq.~(2), one can infer that for an arbitrary number of identical scatterers, $\gamma_+ \leq \gamma_-$. This conclusion seems valid, by arguing that the resonance corresponding to $\omega_-$ has a lower resonance frequency because of the more field overlap with the scatterers, which is also responsible for a stronger scattering loss. However, such intuitive belief is not generally true. Here we consider one extreme example in Fig.~1(b), where identical scatterers have uniformly covered the outer surface of a microresonator except for one vacancy. Assuming the scatterers and the microresonator share the same dielectric constant, the resulting structure can be regarded as a larger-radius microresonator with a negative-dielectric-constant scatterer at the vacancy point. Consequently, the mode that has a nonzero field overlap with the scatterer will incur a blue shift in the resonance frequency (relative to the resonance frequency of the larger-radius microresonator) as well as a linewidth broadening from the scattering. Hence, $\gamma_+ > \gamma_-$, contrary to the result from Eq.~(2). Another problem with the independent-scatterer approach is that it has to track the position of each scatterer, which makes it impractical for problems such as surface roughness caused by imperfect fabrications in high-$Q$ microresonators, where the positions of scatterers are random and only statistical information is available.
\begin{figure}
\includegraphics {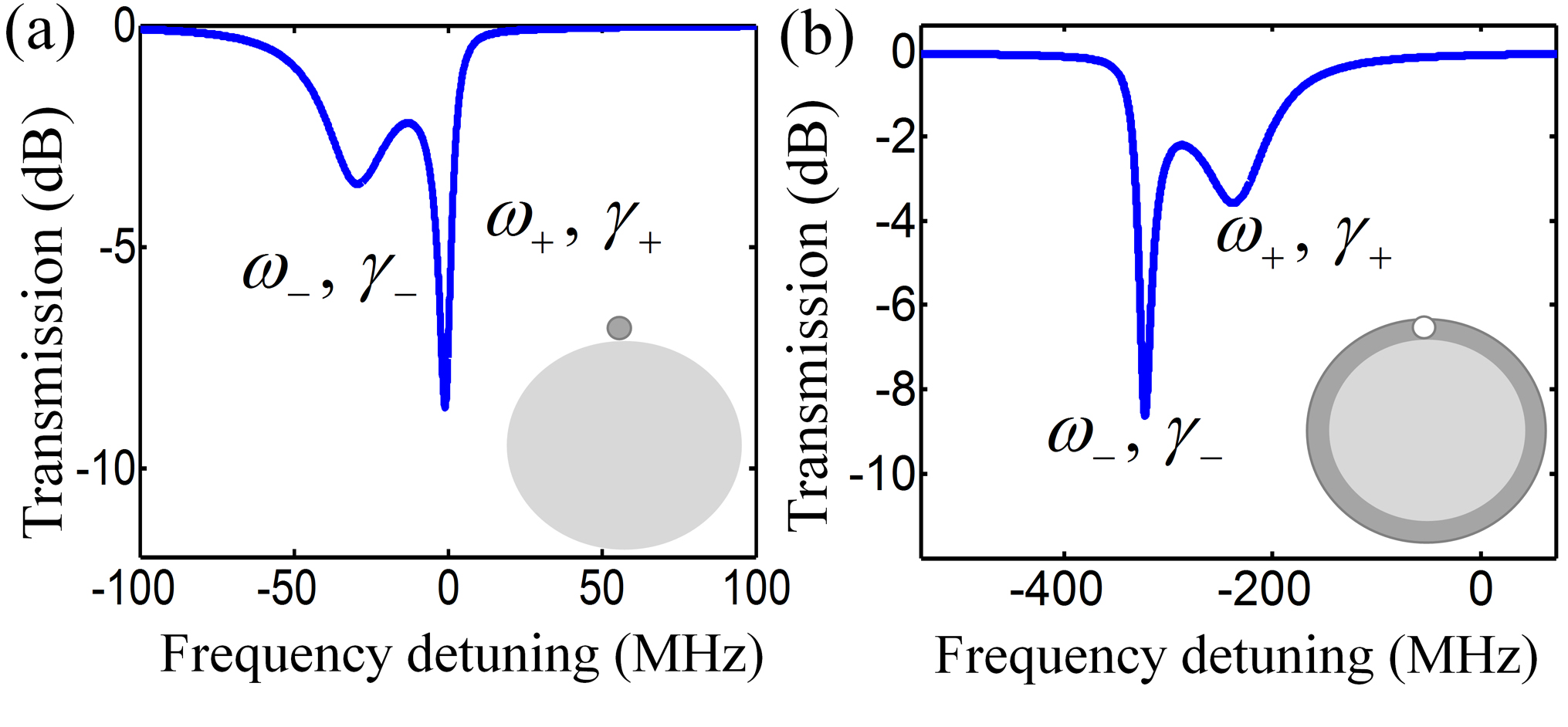}
\caption{(a) Transmission response of a single dielectric scatterer on the surface of a microresonator. (b) Transmission response of numerous dielectric scatterers which uniformly cover the surface of a microresonator except for a vacancy.}
\end{figure}

The second approach, developed mainly for the surface roughness present in high-$Q$ microresonators, employs an intuitive physics model (which is essentially a phenomenological model) to describe the mode splitting and uses the volume current method to obtain the scattering loss \cite{borselli}. Here, we use the microdisk resonator as an example to give an introduction to this approach. For an isolated microdisk resonator (i.e., no external coupling), we have \cite{Haus}
\begin{align}
\frac{da_{\text{ccw}}}{dt} &=\left(-i\omega_c+i\Delta\omega_{\text{ccw}}+  \frac{\gamma_{\text{ccw}}}{2}\right)a_{\text{ccw}} + i\beta_{\text{ccw}}a_{\text{cw}},\\
\frac{da_{\text{cw}}}{dt}&=-\left(i\omega_c+i\Delta\omega_{\text{cw}}+ \frac{\gamma_{\text{cw}}}{2}\right)a_{\text{cw}} + i\beta_{\text{cw}}a_{\text{ccw}},
\end{align}
where $a_{\text{ccw}}$ and $a_{\text{cw}}$ represent the normalized energy amplitudes of the CCW and CW modes, respectively; $\omega_c$ assumes the same meaning as in Eq.~(1), which denotes the unperturbed resonance frequency of the WGM modes; $\Delta\omega_{\text{ccw}}$ and $\Delta\omega_{\text{cw}}$ are the resonance frequency shifts caused by the surface roughness to the CCW and CW modes, respectively; $\gamma_{\text{ccw}}$ and $\gamma_{\text{cw}}$ describe the corresponding scattering loss rates; $\beta_{\text{ccw}}$ is a parameter characterizing the coupling from the CCW to the CW modes and $\beta_{\text{cw}}$ is defined vice versa. From the fact that the CCW and CW modes only differ in their circulating directions, it is expected that
\begin{equation}
\Delta\omega_{\text{ccw}}=\Delta\omega_{\text{cw}},\ \gamma_{\text{ccw}}=\gamma_{\text{cw}},\ \text{and} \ \ \beta_{\text{ccw}}=\beta ^*_{\text{cw}}.
\end{equation}
In fact, from Maxwell's equations, $\Delta\omega_{\text{ccw}}$ and $\beta_{\text{ccw}}$ can be derived as \cite{borselli}
\begin{align}
& \Delta\omega_{\text{ccw}}=-\frac{\omega_c\int\Delta\varepsilon(\bm{r})\left|\bm{E}_{\text{ccw}}(\bm{r})\right|^2\,d^3\bm{r}}
{2\int \varepsilon(\bm{r})\left|\bm{E}_{\text{ccw}}(\bm{r})\right|^2\,d^3\bm{r}},\\
& \beta_{\text{ccw}}=\frac{\omega_c\int\Delta\varepsilon(\bm{r})\bm{E}_{\text{ccw}}^*(\bm{r})\cdot\bm{E}_{\text{cw}}(\bm{r})\,d^3\bm{r}}
{2\int \varepsilon(\bm{r})\left|\bm{E}_{\text{ccw}}(\bm{r})\right|^2\,d^3\bm{r}},
\end{align}
where $\varepsilon(\bm{r})$ and $\Delta\varepsilon(\bm{r})$ correspond to the dielectric constant of a perfect microresonator and that of the surface roughness, respectively, and $\bm{E}_{\text{ccw}}(\bm{r})$ ($\bm{E}_{\text{cw}}(\bm{r})$) is the electric field of the CCW (CW) mode. Similar expressions for  $\Delta\omega_{\text{cw}}$ and $\beta_{\text{cw}}$   exist with the exchange of the CCW and CW subscripts in Eqs.~(6) and (7), and the relations in Eq.~(5) become evident considering $\bm{E}_{\text{ccw}}(\bm{r})$ and $\bm{E}_{\text{cw}}(\bm{r})$ are conjugate to each other (assuming  $\varepsilon(\bm{r})$ is real).

The scattering loss parameter $\gamma_{\text{ccw}}$ ($\gamma_{\text{cw}}$) is obtained using the volume current method by computing the radiation power excited by the polarization current $\bm{J}_{\text{ccw}}(\bm{r})=-i\omega_c\Delta\varepsilon(\bm{r})\bm{E}_{\text{ccw}}(\bm{r})$. For example, the far-field electric field is given by \cite{Kong}
\begin{multline}
\begin{aligned}
\ \bm{E}_{\text{ccw}}^{\text{far}}(\bm{r})=\frac{\omega_c^2e^{ik_0r}}{4\pi\varepsilon_0c^2r} \int &\Delta\epsilon(\bm{r}')\bm{E}_{\text{ccw}}(\bm{r}')\\
&\cdot(1-\hat{\bm r}\hat{\bm r})e^{-ik_0\hat{\bm r}\cdot\bm{r}'}\,d^3\bm{r}',
\end{aligned}
\end{multline}
where $\varepsilon_0$ and $c$ are the permittivity and speed of light of free space, respectively; $k_0$ is the wavenumber; and ($r$, $\theta$, $\phi$) are the spherical coordinates of the far-field position $\bm r$, with ($\hat{\bm r}$, $\hat{\bm \theta}$, $\hat{\bm \phi}$)denoting the orthogonal unit vectors in the directions of increasing ($r$, $\theta$, $\phi$), respectively. In Eq.~(8), we have adopted the $\exp{(-i\omega_c t+ ik_0r)}$ format for the outgoing light to be consistent with the convention used in Eqs.~(3) and (4). The radiation power is calculated by integrating the Poynting vector over the sphere with radius $r$, and $\gamma_{\text{ccw}}$, according to its definition, is given by the power loss rate normalized by the mode energy as
\begin{equation}
\gamma_{\text{ccw}}=\frac{\varepsilon_0 c\iint\left| r \bm{E}_{\text{ccw}}^{\text{far}}(\theta,\phi)\right|^2\sin{\theta}\,d\theta d\phi}
{2\int \varepsilon(\bm{r})\left|\bm{E}_{\text{ccw}}(\bm{r})\right|^2\,d^3\bm{r}}.
\end{equation}
In addition, statistical information of the surface roughness can be inserted into Eq.~(9), which leads to an ensemble average for $\gamma_{\text{ccw}}$ ($\gamma_{\text{cw}}$)\cite{borselli}.

The eigenmodes of the coupled system are then obtained by solving the eigenvalue problem of Eqs.~(3) and (4), which yields
\begin{equation}
a_\pm=\frac{1}{\sqrt{2}} \Bigl( a_{\text{ccw}}\mp \frac{\beta_{\text{ccw}}}{\left|\beta_{\text{ccw}}\right|}a_{\text{cw}} \Bigr),
\end{equation}
where $a_\pm$ denote the two eigenmodes, whose resonance frequencies and scattering loss rates are given by
\begin{align}
\omega_\pm=\omega_c + &\Delta\omega_{\text{ccw}} \pm \left|\beta_{\text{ccw}} \right|,\\
\gamma_\pm &=\gamma_{\text{ccw}}.
\end{align}
Equation (12) predicts that $a_\pm$ should exhibit identical linewidths, while in experiments asymmetric lineshapes are often observed. For practical applications,  $\gamma_\pm$ are generally assumed to be different to fit the model to the experimental data \cite{mohammadoe,borselli}. However, such an assumption contradicts with Eq.~(12). The deficiency of this model can be further illustrated by an intriguing experimental result depicted in Fig.~2, which shows the transmission measurement of a 20-$\mu$m-radius microdisk resonator fabricated on a silicon-on-insulator (SOI) wafer with a 220-nm-thick silicon device layer \cite{qingoe}. By engineering the access waveguide geometry using the pulley coupling scheme \cite{pulley}, only the TE-polarized (electric field parallel to the device layer) first-order radial mode is excited, as confirmed by Fig.~2(a). Among this radial mode family (i.e., resonant modes with the same radial order and different azimuthal orders), four resonances are picked out with zoom-in figures shown in Figs. 2(b)-(e), where significantly different lineshapes are observed. For Fig.~2(b), $\omega_+ \approx \omega_-$, i.e., the mode splitting is negligible; for Figs. 2(c)-(e), doublets appear and we have $\gamma_+ \approx \gamma_-$, $\gamma_+ < \gamma_-$, and $\gamma_+ > \gamma_-$ for each case. It seems difficult to explain the simultaneous occurrence of these features using the intuitive physics model, especially when the corresponding resonances belong to the same radial mode family in one specific microresonator.
\begin{figure}
\includegraphics {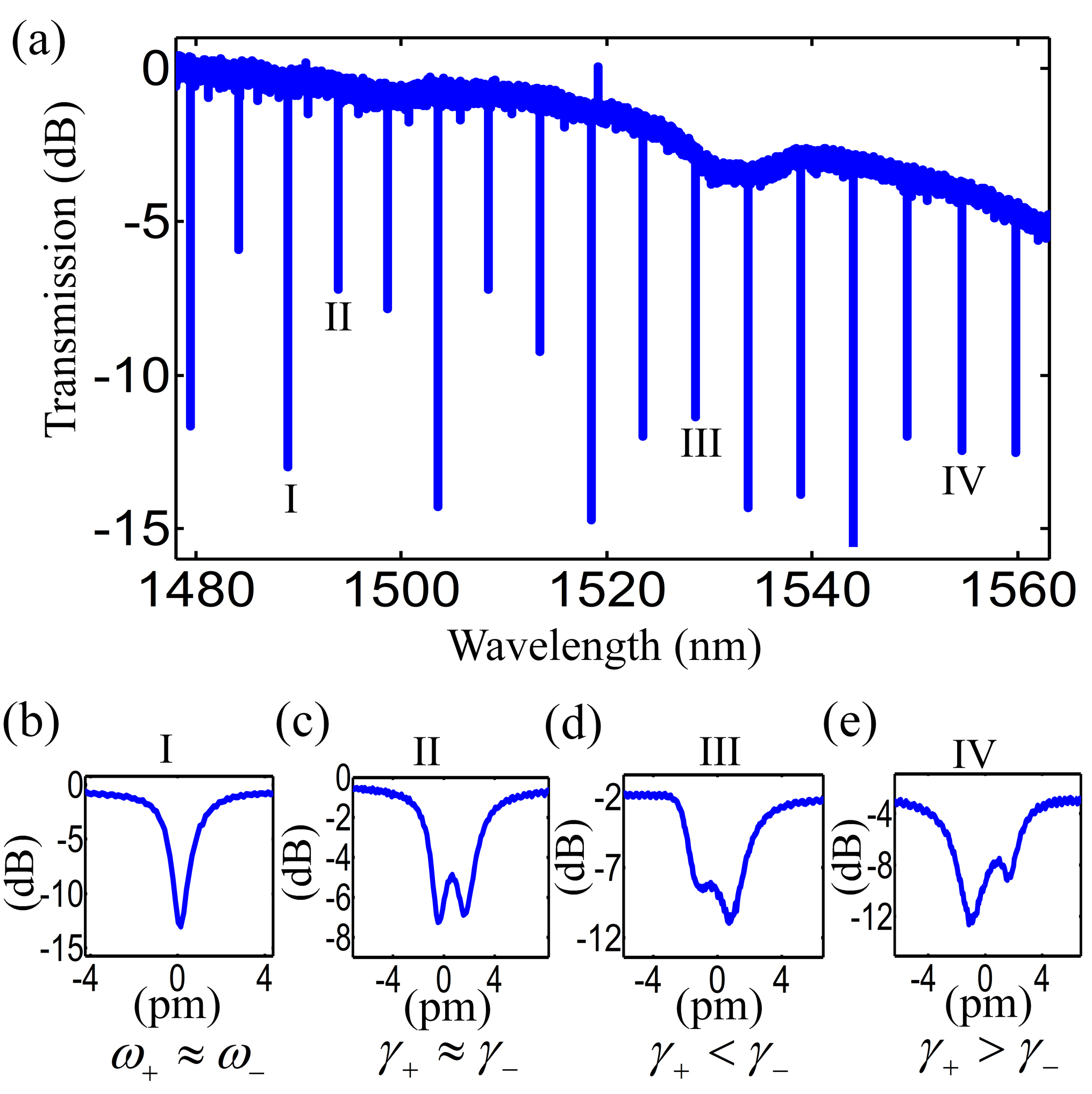}
\caption{ (a) Transmission response of a 20-$\mu$m-radius silicon microdisk which is pulley coupled to a 680-nm-wide access waveguide. Only the fundamental TE mode is phase matched and excited. (b)-(e) Zoom-in figures showing detailed lineshapes of the marked resonances. The resonant modes are over coupled so a higher extinction on resonance indicates a broader linewidth.}
\end{figure}

Summarizing the above discussions, we conclude that the two existing approaches for mode splitting and scattering loss have only achieved partial success. The independent-scatterer approach, based on collective contributions from each individual scatterer, works well when the number of scatterers is small and the scatterers are well separated with each other. On the other hand, the intuitive physics approach, based on a phenomenological model for  the mode splitting and the volume current method for the scattering loss, fits well for  many-scatterer cases such as the surface roughness problem. However, both approaches have difficulties in providing correct scattering loss rates for the eigenmodes of the coupled system. In the independent-scatterer approach, lineshapes are predicted to be symmetric or asymmetric, but $\gamma_+$ is always no more than $\gamma_-$. In the intuitive physics approach, the model cannot predict asymmetric lineshapes (i.e., $\gamma_+ \neq \gamma_-$) in a self-consistent manner.

\section{A unified model}
In this section, we will develop a model that is applicable to an arbitrary number of scatterers attached to the surface of a high-$Q$ microresonator. The approach considers interactions among the CW and CCW modes and the free space continuum, with the coupling provided by scatterers. The derivation here is similar to the independent-scatterer approach as in Refs.~\cite{singleprl,multiple}, but with a key modification that leads to distinct results. Moreover, we show conditions under which our model can be reduced to the results of the two existing approaches discussed in section II in their respective regimes.

For clarity, we use the microdisk resonator as an example for the derivation, while the result is generally applicable to any microresonator with a two-fold degeneracy in its resonance spectrum. For an isolated microdisk resonator (i.e., no external coupling), we can intuitively write down the following equation set:
\begin{gather}
\begin{split}
\frac{da_m}{dt}= -&\left(i\omega_c+\frac{\kappa_0}{2}\right)a_m+ i\sum_{n=1}^{N} \\
                 & \Bigl(\sum_{ m'=\text{cw,ccw}}  g_{n,m,m'}a_{m'} +  \sum_{j}g_{n,m,j}b_j \Bigl),
\end{split}\\
\frac{db_j}{dt}=-i\omega_j b_j + i \sum_{n=1}^{N}\sum_{ m= \text{cw,ccw}}  g_{n,j,m}a_{m},\quad \quad \quad
\end{gather}
where $a_m$ and $b_j$ are the normalized energy amplitudes of the $m$ ($m$ = cw or ccw) WGM and the $j$th free space mode, respectively ($\omega_c$ and $\omega_j$ are their corresponding original resonance frequencies); $\kappa_0$ is the intrinsic cavity loss without including the scattering loss; $g_{n,m,m'}$ is a parameter describing the scattering of the $m$ WGM to the same ($m = m'$) or the counterpropagating ($m \neq m'$) WGM mode induced by the $n$th scatterer; $g_{n,j,m}$ is a similar parameter characterizing the $n$th-scatterer-induced scattering of the $m$ WGM mode to the $j$th free space mode and $g_{n,m,j}$ is defined vice versa. For now we have used a discrete set of eigenmodes $[{b_j}]$ normalized in a finite but large enough volume to represent the free space continuum, and this restriction will be removed later.

In this model, each scatterer is treated as a dipole. The electric field $\bm{E}$ excites the polarization of the $n$th scatterer as $\bm{P}=\varepsilon_0\alpha_n\bm{E}_n$, where $\alpha_n$ and $\bm{E}_n$ are the polarizability and the electric field at the location of the $n$th scatterer, respectively. The interaction of the polarization $\bm P$ with the electric field $\bm{E}$ is given by $-\bm{P}\cdot\bm{E}^*$ \cite{Jackson}, with both the electric fields in  $\bm{P}$ and $\bm{E}$ normalized to their corresponding mode energies. For example, the $j$th free space mode is expressed as
\begin{equation}
\bm{E}_j(\bm{r})=\frac{1}{\sqrt{\varepsilon_0 V_c}}e^{i\bm{k}_j\cdot\bm{r}}\hat{\bm n}_j,
\end{equation}
where $V_c$ is the normalization volume of the free space modes;  $\bm{k}_j$ is the wave vector of the $j$th mode; and $\hat{\bm n}_j$ is the unit polarization vector. Similarly, the energy-normalized electric field of the $m$ WGM has the following form:
\begin{equation}
\bm{E}_m(\bm r)=\frac{f(\bm r)}{\sqrt{\int \varepsilon(\bm r)\left|f(\bm r)\right|^2\, d^3 \bm r}}e^{ik_mx}\hat{\bm n}_m,
\end{equation}
where we have explicitly written out the phase term $\exp{(ik_mx)}$ with $k_m$ being the wavenumber of the $m$ WGM along the mode circulating direction and $x$ being the projection of $\bm r$ on that; $f(\bm r)$ accounts for the amplitude as well as the phase variation other than $\exp{(ik_mx)}$; $\varepsilon(\bm r)$ is the dielectric constant; and $\hat{\bm n}_m$ is the unit vector describing the polarization of the $m$ WGM. To simplify the above expression, we can define a parameter $V_m$ as
\begin{equation}
V_m\equiv \frac{\int \varepsilon(\bm r)\left|f(\bm r)\right|^2\, d^3 \bm r}{\varepsilon_0},
\end{equation}
and $\bm E_m(\bm r)$ can be alternatively expressed as
\begin{equation}
\bm{E}_m(\bm r)=\frac{f(\bm r)}{\sqrt{\varepsilon_0V_m}}e^{ik_mx}\hat{\bm n}_m.
\end{equation}
Note that $V_m$ defined in Eq.~(17) generally does not bear the unit of volume. However, we notice that $f(\bm r)$ is scalable in Eq.~(16). If we normalize the electric field of the WGM mode to that of a reference point, for instance, where the amplitude of the electric field is the maximum, $f(\bm r)$ can be interpreted as the relative field strength and $V_m$ defined above has the unit of volume.

We now proceed to calculate the coupling coefficients $g_{n,m,m'}$, $g_{n,m,j}$, and $g_{n,j,m}$ based on their definitions in Eqs.~(13) and (14). Starting with Maxwell's equations, we have \cite{Kong}
\begin{equation}
\nabla\times(\nabla\times\bm E(\bm r,t)) + \mu\varepsilon (\bm r) \frac{\partial^2 \bm E(\bm r, t)}{\partial t^2} = -\mu \frac{\partial^2 \bm P(\bm r, t)}{\partial t^2},
\end{equation}
where $\mu$ is the permeability of free space. For the $m$ WGM mode,
\begin{equation}
\bm E(\bm r,t)=a_m(t)\bm E_m(\bm r)=e^{-i\omega_ct}(a_m(t)e^{i\omega_ct})\bm E_m(\bm r),
\end{equation}
where we have separated the fast oscillating term $\exp{(-i\omega_ct)}$ with the slowly varying term $a_m(t)\exp{(i\omega_ct)}$. Treating $\bm P(\bm r,t)$ as a first-order perturbation, Eq.~(19) can be approximated as \cite{Kong,Boyd}(also see discussions at the end of Appendix A)
\begin{equation}
2\frac{d}{dt}(a_m(t)e^{i\omega_c t})\approx i \omega_c e^{i\omega_ct}\int \bm P(\bm r,t)\cdot\bm E_m^*(\bm r)\,d^3 \bm r.
\end{equation}
$\bm P(\bm r, t)$ consists of contributions from each scatterer as
\begin{equation}
\begin{split}
\bm P(\bm r,t)=\varepsilon_0 \sum_{n=1}^{N}\alpha_n &\Bigl(\sum_m a_m(t)\bm E_m(\bm r)+ \sum_j b_j(t) \bm E_j(\bm r)\Bigr)\\
                                                   &\times \delta(\bm r -\bm r_n),
\end{split}
\end{equation}
where $\bm r_n$ stands for the position of the $n$th scatterer. Substituting the detailed expression of $\bm P(\bm r,t)$ into Eq.~(21) and comparing it to Eq.~(13), we arrive at
\begin{align}
g_{n, m,m'}& = \frac{\alpha_n\omega_c\left|f(\bm r_n)\right|^2}{2V_m}e^{i(k_{m'}-k_m)x_n}, \\
g_{n,m,j}=&\frac{\alpha_n \omega_c f^*(\bm r_n)}{2\sqrt{V_m V_c}}e^{i(\bm k_j\cdot \bm r_n-k_m x_n)}\Bigl (\hat{\bm n}_j \cdot \hat{\bm n}_m(\bm r_n) \Bigr ),
\end{align}
where $x_n$ is the projection of the $n$th-scatterer's position $\bm r_n$ along the WGM circulating direction. In Eq.~(24), the dependence of the polarization of the $m$ WGM mode $\hat{\bm n}_m$ on the position of the $n$th scatterer $\bm r_n$ has been explicitly expressed, given that $\hat{\bm n}_m$ is not necessarily a constant vector (for example, the dominant electric field for the TE-polarized WGM is $E_\phi$, which is in the direction of $\hat{\bm \phi}$). In deriving $g_{n,m,j}$, we have used the fact that only those free space modes that can resonate with the WGM modes (i.e., $\omega_j \approx \omega_c$) need to be considered, which will be proven soon. Taking a similar procedure for Eq.~(14) leads us to
\begin{equation}
g_{n,j,m}=\frac{\alpha_n\omega_c f(\bm r_n)}{2\sqrt{V_m V_c}}e^{i(-\bm k_j\cdot \bm r_n+k_m x_n)}\Bigl (\hat{\bm n}_j \cdot \hat{\bm n}_m(\bm r_n) \Bigr).
\end{equation}
Note that $g_{n,m,j}$ and $g_{n,j,m}$ obtained here are different from those obtained in Refs.~\cite{singleprl,multiple}, where the derivation is based on the interaction among quantized fields and the $\exp(i\bm k_j\cdot \bm r_n)$ ($\exp(-i\bm k_j\cdot \bm r_n)$) factor is missing in $g_{n,m,j}$ ($g_{n,j,m}$).

With the knowledge of the coupling coefficients $g_{n,m,m'}$, $g_{n,m,j}$, and $g_{n,j,m}$, we are ready to solve Eqs.~(13) and (14). Instead of studying the fast oscillating terms $a_m(t)$ and $b_j(t)$, it is more convenient to work with the slowly changing variables $\bar{a}_m(t) \equiv a_m(t)\exp{(i\omega_ct)}$ and
$\bar{b}_j(t) \equiv b_j(t)\exp{(i\omega_jt)}$. We solve $\bar{b}_j(t)$ in Eq.~(14) in terms of $\bar{a}_m(t)$ and insert it back into Eq.~(13). After some arrangements, we obtain
\begin{equation}
\begin{aligned}
\frac{d\bar a_m(t)}{dt}= &-  \frac{\kappa_0}{2}  \bar a_m(t)  +  i\sum_{n=1}^{N}\sum_{m'=\text{cw,ccw}}g_{n,m,m'}\bar a_{m'}(t)\\
&\ \ -\sum_{m'=\text{cw,ccw}} \sum_{n=1}^{N}\sum_{n'=1}^{N}\sum_{j} g_{n,m,j}g_{n',j,m'}\\
     &\quad \quad \quad \times\int\limits_{-\infty}^{t} e^{i(\omega_j-\omega_c)(t'-t)}\bar a_{m'}(t')\, dt'.
\end{aligned}
\end{equation}
Because the free space modes have a high mode density, we can replace the summation over the mode number $j$ by an integral over the wave vector space as \cite{Kong}
\begin{equation}
\sum_j \longrightarrow \frac{V_c}{(2\pi)^3} \sum_{\hat{\bm       n}_k}\int\limits_{-\pi}^{\pi}\,d\phi\int\limits_{0}^{\pi}\sin \theta \,d\theta \int\limits_{0}^{\infty} k^2\,dk,
\end{equation}
where ($k$, $\theta$, $\phi$) are the spherical coordinates of the wave vector $\bm k$, with ($\hat{\bm k}$, $\hat{\bm \theta}$, $\hat{\bm \phi}$) denoting the orthogonal unit vectors in the directions of increasing ($k$, $\theta$, $\phi$), respectively. $\hat{\bm n}_k$ describes two possible polarizations corresponding to $\hat{\bm k}$, and hence can be in the direction of $\hat{\bm \theta}$ or $\hat{\bm \phi}$.

Inserting the expressions of $g_{n,m,j}$ and $g_{n,j,m}$ into the last term on the right side of Eq.~(26), we arrive at
\begin{equation}
\begin{split}
\sum_{m'=\text{cw,ccw}}\sum_{n=1}^{N}\sum_{n'=1}^{N}\frac{\alpha_n \alpha_{n'}\omega_c^2 f^*(\bm r_n) f(\bm r_{n'})}{4(2\pi)^3 V_m c^3} e^{i(k_{m'} x_{n'}-k_m x_n)}\\
\times\iint \sum_{\hat{\bm n}=\hat{\bm \theta},\hat{\bm \phi}} \Bigl( \hat{\bm n}\cdot \hat{\bm n }_m(\bm r_n)\Bigr)\Bigl( \hat{\bm n}\cdot \hat{\bm n }_{m'}(\bm r_{n'})\Bigr) \sin \theta \,d\theta d \phi\\
\times \int\limits_{0}^{\infty} \omega^2 e^{i \frac{\omega}{c}\hat{\bm k} \cdot (\bm r_n - \bm r_{n'})} \, d\omega \int\limits_{-\infty}^{t} e^{i(\omega-\omega_c)(t'-t)}\bar a_{m'}(t')\, dt',
\end{split}
\end{equation}
where we have used the relation $\omega=kc$ to replace $k$ in the integral in Eq.~(27).  In the integration over $t'$, because the microresonator has a high $Q$, it is reasonable to assume $\bar{a}_m(t')$ varies sufficiently slowly over a few optical cycles so that it can be evaluated at the $t = t'$ \cite{singleprl}. In consequence,
\begin{equation}
\int\limits_{-\infty}^{t} e^{i(\omega-\omega_c)(t'-t)}\bar a_{m'}(t')\, dt' \approx \pi \delta(\omega-\omega_c) \bar a_{m'}(t),
\end{equation}
which indicates only the free space modes with resonance frequencies around $\omega_c$ need to be considered. In Eq.~(28), another simplification can be carried out for the summation over the polarizations of the free space modes with the help of the following vector identity:
\begin{equation}
\sum_{\hat{\bm n}=\hat{\bm \theta},\hat{\bm \phi}} (\hat{\bm n} \cdot \bm a) (\hat{\bm n} \cdot \bm b) =\left[ (1-\hat{\bm k}\hat{\bm k}) \cdot \bm a \right] \left[ (1-\hat{\bm k}\hat{\bm k}) \cdot \bm b \right],
\end{equation}
which can be easily verified for arbitrary vectors $\bm a$ and $\bm b$.

As a result, we have
\begin{equation}
\frac{d\bar a_m(t)}{dt}= -  \frac{\kappa_0}{2}  \bar a_m(t) +   \sum_{m'=\text{cw,ccw}} (iG_{m,m'} - \frac{\Gamma_{m,m'}}{2})\bar a_{m'}(t),
\end{equation}
with
\begin{gather}
G_{m,m'} \equiv \frac{\alpha_n\omega_c\left|f(\bm r_n)\right|^2}{2V_m}e^{i(k_{m'}-k_m)x_n},\quad \quad \quad \quad \quad \quad \quad \quad \quad  \\
\begin{split}
\Gamma_{m,m'} \equiv & \sum_{n=1}^{N}\sum_{n'=1}^{N}\frac{\alpha_n \alpha_{n'}\omega_c^4 f^*(\bm r_n) f(\bm r_{n'})}{(4\pi)^2 V_m c^3} e^{i(-k_m x_n +k_{m'} x_{n'})}\\
 \iint & \left[ (1-\hat{\bm k}\hat{\bm k}) \cdot \hat{\bm n}_m(\bm r_n) \right] \cdot \left[ (1-\hat{\bm k}\hat{\bm k}) \cdot \hat{\bm n}_{m'}(\bm r_{n'}) \right]\\
& \times e^{i k_0 \hat{\bm k} \cdot (\bm r_n - \bm r_{n'})} \sin \theta \, d\theta d\phi,
\end{split}
\end{gather}
where $k_0 = \omega_c /c$ is the wavenumber of light with angular frequency $\omega_c$ in free space. The integral in $\Gamma_{m,m'}$ involves an integration over the spherical surface in the wave vector space, and its value depends on the polarization of the WGM modes as well as the relative positions of scatterers. Hence, it is a geometric factor.

\subsection{Comparison with the independent-scatterer approach}
The geometric integral in $\Gamma_{m,m'}$ given by Eq.~(33) can be computed  for any given $\bm r_n$ and $\bm r_{n'}$. For example, if the WGM mode is TM-polarized (magnetic field parallel to the device layer), $ \hat{\bm n}_m = \hat{\bm z}$ \ (see Fig.~3(a)). In addition, we can choose the $x$ axis to be in the direction of $\bm r_n -\bm r_{n'}$ . The geometric integral in Eq.~(33) can then be simplified as
\begin{equation}
\begin{split}
&\int\limits_{0}^{\pi} \sin^3{\theta} \, d \theta \int\limits_{-\pi}^{\pi} e^{ik_0 \left |\bm r_n -\bm r_{n'} \right| \sin{\theta} \cos{\phi}} \, d\phi \\
&= 2\pi \int\limits_{0}^{\pi} \sin^3{\theta} J_0(k_0 d_{n,n'}\sin \theta) \, d \theta = \frac{8\pi}{3}p(k_0 d_{n,n'}),
\end{split}
\end{equation}
where $d_{n,n'} \equiv \left| \bm r_n -\bm r_{n'} \right|$; $J_0(x)$ is the Bessel function of the first kind of order zero; and $p(x)$ is defined as
\begin{equation}
p(x) \equiv \frac{3}{4} \int\limits_{0}^{\pi} \sin^3{\theta} J_0(x\sin \theta) \, d \theta.
\end{equation}
In deriving Eq.~(34), integral representations of the Bessel functions are used and mathematical details are left to Appendix A. From Eq.~(34), we find that the geometric integral in $\Gamma_{m,m'}$ is only a function of the separation distance between scatterers; consequently,
\begin{equation}
\begin{split}
\Gamma_{m,m'}= \sum_{n=1}^{N}\sum_{n'=1}^{N} & \frac{\alpha_n \alpha_{n'}\omega_c^4 f^*(\bm r_n) f(\bm r_{n'})}{6 \pi V_m c^3} \\
&  \times e^{i(-k_m x_n +k_{m'}x_{n'})} p(k_0 d_{n,n'}).
\end{split}
\end{equation}

\begin{figure}
\includegraphics {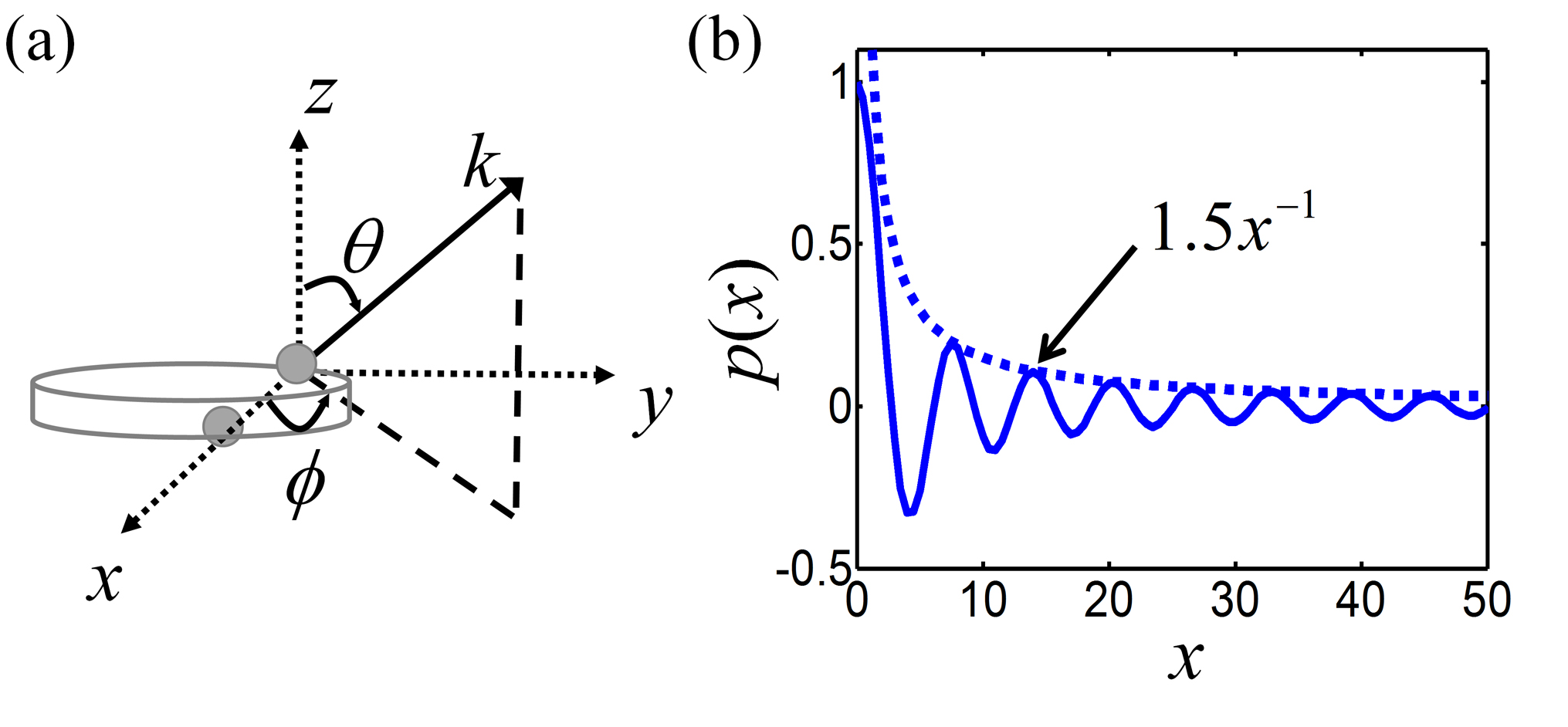}
\caption{ (a) Illustration of the adopted coordinate system for the calculation of the geometric integral in Eq.~(34): the $z$ axis is perpendicular to the microresonator, and the $x$ axis is chosen to be along the relative position of the two scatterers under consideration. (b) The solid line is the numerical result of $p(x)$ defined in Eq.~(35), and the dotted line is $3/2x$, which corresponds to the envelop of $p(x)$ when $x$ is large. The exact expression of $p(x)$ is given by Eq.~(A7) in Appendix A.}
\end{figure}

\noindent In Fig.~3(b), numerical values of $p(x)$ are evaluated. For $x = 0$, $p(0) =1$; when the argument $x$ is large, the envelope of $p(x)$ decreases at a rate of $x^{-1}$, which can be proved from a rigorous calculation (see Appendix A). In the independent-scatterer approach discussed in section II, only the $n = n'$ terms  (i.e., $d_{n,n'} = 0$ and $p(0)=1$) are considered in the double summation over $n$ and $n'$ for $\Gamma_{m,m'}$, and the $n \neq n'$ terms are neglected, with the hope that the contribution from these terms is small if scatterers are well separated with each other \cite{multiple}. From Fig.~3(b), we estimate that a reasonably large separation should be on the order of $d_{n,n'}/\lambda_0> 2.4$ ($|p(k_0d_{n,n'})|< 0.1$), with $\lambda_ 0$ corresponding to the free space wavelength ($\lambda_0=2 \pi c/\omega_c$). Otherwise, the omission of $n \neq n'$ terms can introduce significant errors and even lead to erroneous conclusions (such as $\gamma_+ \leq \gamma_-$, see discussions in section II). In section IV, we will examine one numerical example for two scatterers attached to the surface of a microresonator, where we show it is essential to include $p(x)$ for a complete understanding of the simulation results.

\subsection{Comparison with the intuitive physics approach}
When the number of scatterers is large, the forms of $\Gamma_{m,m'}$ given by Eq.~(33) (or Eq.~(36)) is  not that convenient to work with. From Eq.~(33), we notice that if we define
\begin{equation}
\begin{split}
\ \bm{S}_m(\theta, \phi) \equiv \frac{\omega_c^2}{4\pi \sqrt{V_m c^3}} \sum_{n=1}^{N} & \alpha_n f(\bm r_n)e^{ik_m x_n}\hat{\bm n}_m(\bm r_n) \\
& \cdot(1-\hat{\bm k}\hat{\bm k})e^{-ik_0\hat{\bm k} \cdot \bm r_n},
\end{split}
\end{equation}
$\Gamma_{m,m'}$ can be rewritten as
\begin{equation}
\Gamma_{m,m'} =\iint \bm S^*_m(\theta,\phi) \cdot \bm S_{m'}(\theta,\phi) \sin \theta \, d\theta d\phi.
\end{equation}
The expression of $\bm S_m(\theta, \phi)$ in Eq.~(37) is invariant if we scale $f(\bm r)$; hence, we can remove the restriction of $\bm E_m(\bm r)$ defined in Eq.~(16) (or Eq.~(18)) which requires it to be energy normalized, and extend the definition of $\bm S_m(\theta, \phi)$ to arbitrary $\bm E_m(\bm r)$ as
\begin{equation}
\begin{split}
\bm{S}_m(\theta, \phi)=\frac{\omega_c^2}{4\pi} \sqrt{\frac{\varepsilon_0}{U_m c^3}} \sum_{n=1}^{N} & \alpha_n \bm E_m(\bm r_n) \\
      &\cdot (1-\hat{\bm k}\hat{\bm k})e^{-ik_0\hat{\bm k} \cdot \bm r_n},
\end{split}
\end{equation}
with $U_m$ defined as
\begin{equation}
 U_m \equiv \int \varepsilon(\bm{r})\left|\bm{E}_m(\bm{r})\right|^2\,d^3\bm{r},
\end{equation}
which corresponds to the energy of the $m$ WGM mode. Also, any distribution of scatterers on the surface of a microresonator can be treated as a special case of surface roughness, with the dielectric perturbation given by
\begin{equation}
\Delta \varepsilon(\bm r)=\sum_{n=1}^{N} \varepsilon_0 \alpha_n \delta(\bm r- \bm r_n).
\end{equation}
Combining Eqs.~(39) and (41), a general form for $\bm S_m(\theta, \phi)$ is found as
\begin{equation}
\begin{split}
\bm{S}_m(\theta, \phi)=\frac{\omega_c^2}{4\pi} \sqrt{\frac{\varepsilon_0}{U_m c^3}} \int & \Delta \varepsilon (\bm r) \bm E_m(\bm r) \\
                     &\cdot (1-\hat{\bm k}\hat{\bm k})e^{-ik_0\hat{\bm k} \cdot \bm r} \, d^3 \bm r.
\end{split}
\end{equation}
A direct comparison shows that, except a constant,  $\bm S_m(\theta, \phi)$ given by Eq.~(42) in our model is equal to $r\bm{E}_{\text{ccw}}^{\text{far}}$  with $\bm{E}_{\text{ccw}}^{\text{far}}$ given by Eq.~(8) in the volume current method, if we identify ($\hat{\bm k}$, $\hat{\bm \theta}$, $\hat{\bm \phi}$) (which are the coordinates of the wave vector $\bm k$) in Eq.~(42) with ($\hat{\bm r}$, $\hat{\bm \theta}$, $\hat{\bm \phi}$)(which are the coordinates of the far-field position $ \bm r$)  in Eq.~(8). Furthermore, according to Eq.~(31), $\Gamma_{m,m}$ describes the scattering loss rate of the CCW (CW) WGM mode, similar to $\gamma_\text{ccw}$ in the phenomenological model in Eq.~(3). Their difference is that $\Gamma_{m,m}$, given by Eq.~(38), is a spherical integration of $|\bm S_m(\theta, \phi)|^2$ in the wave vector space, and $\gamma_\text{ccw}$, given by Eq.~(9), is a spherical integration of $|r \bm{E}_{\text{ccw}}^{\text{far}}|^2$  in the coordinate space. From the mathematical point of view, however, there is no difference in their detailed expressions and one can verify $\gamma_{\text{ccw}}=\Gamma_{m,m}$.

The analogy between our approach and the intuitive physics approach discussed in section II can be carried on further. With the help of Eq.~(41), $G_{m,m'}$  defined in Eq.~(32) can also be extended to the general case of surface roughness as
\begin{equation}
G_{m,m'}=\frac{\omega_c}{2U_m}\int \Delta\varepsilon (\bm r)\bm E^*_m(\bm r)\cdot \bm E_{m'} (\bm r) \, d^3 \bm r.
\end{equation}
From here on, we will use a slightly different notation for the ease of comparison, that is, we use the subscripts $m$ ($> 0$) and $-m$ ($< 0$) to stand for the CCW and the CW WGM modes with the azimuthal order $m$, respectively. Comparing Eq.~(43) to Eqs.~(6) and (7), it is easy to recognize $\Delta\omega_{\text{ccw}}=-G_{m,m}$ and $\beta_{\text{ccw}}=G_{m,-m}$. Adding the already known relation $\gamma_{\text{ccw}}=\Gamma_{m,m}$, the key difference between our model given by Eq.~(31) and the phenomenological model given by Eqs.~(3) and (4) is that we have an additional coupling coefficient $\Gamma_{m,-m}$ in Eq.~(31). After converting $\bar a_m(t)$ back to $a_m(t)$ and setting $\kappa_0$ to be zero (meaning only scattering loss is considered), Eq.~(31) can be explicitly expressed as
\begin{align}
&\frac{da_m}{dt} = - \left(i\omega_c -iG_{m,m}+  \frac{\Gamma_{m,m}}{2} \right)a_m \notag \\
                   &\qquad \qquad\qquad \qquad+\left(iG_{m,-m} - \frac{\Gamma_{m,-m}}{2} \right)a_{-m},\\
&\frac{da_{-m}}{dt} = -\left(i\omega_c -iG_{m,m}+  \frac{\Gamma_{m,m}}{2}\right)a_{-m} \notag \\
                     & \qquad \qquad \qquad \qquad +\left(iG^*_{m,-m} - \frac{\Gamma^*_{m,-m}}{2}\right)a_{m},
\end{align}
where we have used the facts $G_{-m,-m}=G_{m,m}$,\ $G_{-m,m}=G^*_{m,-m}$, \ and $\Gamma_{-m,m}=\Gamma^*_{m,-m}$, all of which are evident from their definitions (assuming $\Delta \varepsilon(\bm r)$ is real). We also use the relation $\Gamma_{-m,-m}=\Gamma_{m,m}$, which is not that obvious from Eq.~(38), since from the expression of  $\bm S_m(\theta, \phi)$ given by Eq.~(42),  $\bm S_m(\theta, \phi)$ and  $\bm S_{-m}(\theta, \phi)$  are not conjugate to each other. However, at current stage, we assume  $\Gamma_{-m,-m}=\Gamma_{m,m}$ simply from the fact that the CW and CCW modes have the same scattering loss rate, and we will prove that in the next subsection.

Solving the eigenmodes of Eqs.~(44) and (45) yields
\begin{align}
a_\pm =\frac{1}{\sqrt{1+ \left|\eta \right|^2}} \bigl( a_{\text{ccw}} \mp \eta a_{\text{cw}} \bigr), \quad \quad  \quad  \\
\omega_\pm =\omega_c - G_{m,m} \pm \text{Re}\,\bigl(\eta(G^*_{m,-m} + \frac{i}{2} \Gamma^*_{m,-m})\bigr),\\
\gamma_\pm =\Gamma_{m,m} \mp 2 \text{Im} \, \bigl(\eta(G^*_{m,-m} + \frac{i}{2} \Gamma^*_{m,-m})\bigr), \quad
\end{align}
where
\begin{equation}
\eta= \pm \sqrt{\frac{G_{m,-m} +  \frac{i}{2} \Gamma_{m,-m}}{G^*_{m,-m} +  \frac{i}{2} \Gamma^*_{m,-m}}}.
\end{equation}
The sign of $\eta$ is chosen to ensure that $\omega_+$ defined in Eq.~(47) is no less than $\omega_-$. From Eq.~(49), one immediate observation is that in general the CW and CCW components in the eigenmodes are not of equal weight (i.e., $|\eta|$ is not necessarily equal to 1), in contrast to the result from Eq.~(10), where the CW and CCW modes are equally weighted. Furthermore, from Eqs.~(47)-(49), we have
\begin{align}
&(\omega_\pm-\omega_c + G_{m,m})  (\gamma_\pm  - \Gamma_{m,m}) = \notag \\
&\qquad \qquad \qquad \qquad  \qquad -\text{Re}\,(G^*_{m,-m}\Gamma_{m,-m}),\\
&(\omega_\pm-\omega_c + G_{m,m})^2 -  (\gamma_\pm-\Gamma_{m,m})^2/4= \notag \  \\
&\qquad \qquad \qquad \qquad  \qquad  |G_{m,-m}|^2-|\Gamma_{m,-m}|^2/4.
\end{align}
For $\omega_+$, the first multiplying factor on the left side of Eq.~(50) is no less than zero, and we conclude that $\gamma_+ <\gamma_-$ if $\text{Re}\,(G^*_{m,-m}\Gamma_{m,-m}) >0$ and $\gamma_+ >\gamma_-$  if $\text{Re}\,(G^*_{m,-m}\Gamma_{m,-m}) <0$. The case that $\text{Re}\,(G^*_{m,-m}\Gamma_{m,-m}) =0$ is interesting, which corresponds to two possibilities, the first being the two eigenmodes have the same linewidth but different resonance frequencies (i.e.,\ $\omega_+ > \omega_-$, $\gamma_+ =\gamma_-=\Gamma_{m,m}$) and the second being the two eigenmodes have the same resonance frequency but different linewidths (i.e.,\ $\omega_+ = \omega_-= \omega_c-G_{m,m}$, $\gamma_+ \neq \gamma_-$), depending on whether $|G_{m,-m}|>|\Gamma_{m,-m}|/2$ or $|G_{m,-m}|<|\Gamma_{m,-m}|/2$  (see Eq.~(51)). One trivial condition for
$\text{Re}\,(G^*_{m,-m}\Gamma_{m,-m}) =0$ is $G_{m,-m}=0$ , but since $\Gamma_{m,-m}$ generally is nonzero, the two resonances would have different scattering loss rates and can manifest themselves under different excitation conditions (i.e., weak and strong couplings). Thus, the two eigenmodes are not degenerate in the strict sense. However, both the experiment in Ref.~\cite{double} and the numerical study in Ref.~\cite{multiple} fail to observe the distinction of the scattering loss rates between the two eigenmodes when they overlap in the resonance frequency, mostly because only one transmission result with one particular excitation is available. In section IV, we will show numerical examples that clearly demonstrate when \ $\omega_+ = \omega_-$,  $\gamma_+$ generally is not equal to $\gamma_-$.

Next, we would like to see how the results given by Eqs.~(46)-(48) can be reduced to those of the phenomenological model derived in Eqs.~(10)-(12). For dielectric perturbations  in high-$Q$ microresonators, usually (but not always) $|G_{m,-m}|\gg | \Gamma_{m,-m}|/2$. Under this condition, $\eta$ given by Eq.~(49) can be approximated as $\sqrt{G_{m,-m}/G^*_{m,-m}}$ , and Eqs.~(46)-(48) are simplified as
\begin{align}
a_\pm \approx & \ \frac{1}{\sqrt 2}  \Bigl(  a_{\text{ccw}} \mp \frac{G_{m,-m}}{|G_{m,-m}|} a_{\text{cw}}  \Bigr), \\
\omega_\pm &\approx \omega_c - G_{m,m} \pm |G_{m,-m}|, \\
\gamma_\pm \approx & \  \Gamma_{m,m} \mp \frac{\text{Re}\, (G ^*_{m,-m} \Gamma_{m,-m})}{|G_{m,-m}|}.
\end{align}
Comparing Eqs.~(52)-(54) to Eqs.~(10)-(12), with the equalities of $\Delta\omega_{\text{ccw}}=-G_{m,m}$  and $\beta_{\text{ccw}}=G_{m,-m}$ (see discussions following Eq.~(43)), we find only $\gamma_\pm$ are different between the two approaches. In fact, there is a simple explanation for the result of Eq.~(54). The electric fields corresponding to the eigenmodes $a_\pm$ are the eigenvectors of Eqs.~(44) and (45), which can be solved as
\begin{equation}
\bm{E}_\pm (\bm r) \approx \frac{1}{\sqrt 2} \Bigl( \frac{G_{m,-m}}{|G_{m,-m}|} \bm E_m (\bm r) \mp \bm E_{-m} (\bm r) \Bigr).
\end{equation}
The associated scattering loss rate can then be calculated based on the volume current method. The far-field electric fields corresponding to $\bm{E}_\pm (\bm r)$ have the same linear combinations as in Eq.~(55) by those of the CW and CCW modes. Thus,
\begin{equation}
\left| \bm{E}^{\text{far}}_\pm (\bm r) \right|^2 \approx \left |\bm{E}^{\text{far}}_m (\bm r) \right|^2 \mp \frac{\text{Re}\, (G ^*_{m,-m}\bm{E}^{\text{far}*}_m (\bm r) \bm{E}^{\text{far}}_{-m} (\bm r) )}{|G_{m,-m}|}.
\end{equation}
The scattering loss rate involves an integration of  $\left| \bm{E}^{\text{far}}_\pm (\bm r) \right|^2$ over the sphere with radius of $r$ as in Eq.~(9). Using the equivalence we have established between  $\bm S_m(\theta, \phi)$ and  $r\bm{E}_m^{\text{far}}$ (see discussions following Eq.~(42)) as well as the expression of $\Gamma_{m,m'}$ given by Eq.~(38), we find the spherical integration of $\left |\bm{E}^{\text{far}}_m (\bm r) \right|^2$ is equal to $\Gamma_{m,m}$ and the spherical integration of $\bm{E}^{\text{far}*}_m (\bm r) \bm{E}^{\text{far}}_{-m} (\bm r)$ is equal to $\Gamma_{m,-m}$, and Eq.~(54) becomes apparent.

\subsection{Formulation in the Fourier domain}
In this section, we will derive a useful formulation for the calculation of parameters needed to obtain mode splitting and scattering loss, i.e., $G_{m,m}$, $G_{m,-m}$, $\Gamma_{m,m}$ , and $\Gamma_{m,-m}$ in Eqs.~(44) and (45). We still use the microdisk resonator as an example, and we further assume that the scatterers are uniform along the microdisk slab thickness (this contains the Rayleigh-scatterer case, for which the detailed shape of scatterers is not important). As a result, the surface roughness $\Delta\varepsilon(\bm r)$ defined in Eq.~(41) can be assumed to have the following form:
\begin{equation}
\Delta\varepsilon(\bm r)= \varepsilon_0  \delta(r-R) \Delta\varepsilon_r(\phi) \text{rect}(z/h),
\end{equation}
where  $R$ is the radius of the disk; $\phi$ is the azimuth, as shown in Fig.~3(a); $\Delta\varepsilon_r(\phi)$ characterizes the dielectric perturbations along the periphery of the microdisk resonator; $h$ is the slab thickness; and $\text{rect}(x)$ stands for the rectangular function \cite{math}. Because of the inherent periodic boundary condition,   $\Delta\varepsilon_r(\phi)$ can be expanded in terms of periodic harmonics along the microdisk periphery as \cite{qingol}
\begin{equation}
\Delta\varepsilon_r(\phi)= \frac{1}{2\pi} \sum_{n} F(k_n)e^{in\phi},
\end{equation}
where $F(k_n)$ is the Fourier component of $\Delta\varepsilon_r(\phi)$ with $k_n = n/R$ ($n = 0, \,\pm 1,\,\pm 2,\dots$). The Fourier transform of Eq.~(58) gives
\begin{equation}
F(k_n)= \int\limits_{0}^{2\pi}\Delta\varepsilon_r(\phi)e^{-in\phi}\,d\phi.
\end{equation}
For dielectric perturbations,\, $\Delta\varepsilon_r(\phi)$ is real, which yields $F(k_n) = F^*(k_{-n})$. Inserting the form of $\Delta\varepsilon_r(\phi)$ given by Eq.~(58) into Eq.~(43), $G_{m,m}$ and $G_{m,-m}$  are obtained as
\begin{align}
&G_{m,m}=g_0 F(k_0),\\
&G_{m,-m}=g_0F(k_{2m}),
\end{align}
with
\begin{equation}
g_0 \equiv \frac{\varepsilon_0\omega_cRh}{2U_m}\left|\bar{\bm E}_m(R,0)\right|^2,
\end{equation}
where $\left|\bar{\bm E}_m(R,0)\right|$ is the amplitude of the electric field $\bm E_m(\bm r)$  at the surface ($r = R$, $\phi= 0$) after averaging along the slab thickness.

To obtain  $\Gamma_{m,m}$ and $\Gamma_{m,-m}$, we first calculate $\bm S_m(\theta,\phi)$   based on Eq.~(42). For the TM-polarized WGM modes, $ \hat{\bm n}_m = \hat{\bm z}$ , and Eq.~(42) can be simplified as \cite{qingol}
\begin{align}
\bm S_m(\theta, \phi)& =\sqrt{C_0}(-\sin \theta \hat{\bm \theta})\frac{1}{2\pi}\sum_n F^*(k_n) \notag \\
                           &\quad \quad \times \int\limits_{-\pi}^{\pi}e^{i(m-n)\phi' -ik_0R\sin \theta \cos{(\phi'- \phi)}}\, d\phi' \notag \\
                     =\sqrt{C_0}&(-\sin \theta \hat{\bm \theta})\sum_n F^*(k_{m+n})(ie^{i\phi})^{-n}J_n(k_0R\sin \theta),
\end{align}
where the integral representation of $J_n(x)$ is used (see Appendix A) and  $C_0$ is defined as
\begin{equation}
C_0 \equiv \frac{\varepsilon_0 w_c^4 R^2 h^2}{16 \pi^2 U_m c^3} \left|\bar{\bm E}_m(R,0)\right|^2.
\end {equation}
Likewise, we have
\begin{align}
\bm S_{-m}(\theta, \phi)=\sqrt{C_0}(-\sin \theta \hat{\bm \theta})\sum_n & F(k_{m+n})(ie^{-i\phi})^{-n} \notag \\
                                            &\times J_n(k_0R\sin \theta).
\end{align}
Substituting $\bm S_m(\theta,\phi)$ and $\bm S_{-m}(\theta,\phi)$ into Eq.~(38),  $\Gamma_{m,m}$ and $\Gamma_{m,-m}$ are found as
\begin{align}
 & \Gamma_{m,m}= 2\pi C_0 \sum_n \left| F(k_{m+n}) \right|^2  \notag \\
  & \qquad \qquad \qquad \qquad   \times \int\limits_{0}^{\pi}J^2_n(k_0R\sin \theta)\sin^3\theta \, d\theta,
\end{align}
and
\begin{align}
& \Gamma_{m,-m}= 2\pi C_0 \sum_n  F(k_{m+n}) F(k_{m-n})  \notag \\
& \qquad \qquad \qquad \qquad       \times \int\limits_{0}^{\pi}J^2_n(k_0R\sin \theta)\sin^3\theta \, d\theta .
\end{align}
 One can verify that if we replace $m$ by $-m$ in Eq.~(66), after some arrangements (replacing the summation index $n$ by $-n$ and using $F(k_{-n}) = F^*(k_{n})$, $J_{-n}(x) = (-1)^n J_n(x)$), it will lead to the same expression. This proves $\Gamma_{m,m}=\Gamma_{-m,-m}$ , which we have already used in Eqs.~(44) and (45).

For the TE-polarized WGM modes, the dominant electric field is $E_\phi$, which is in the direction of $\hat{\bm \phi}$. Equation (42) then becomes \cite{qingol}
\begin{align}
\bm S_m(\theta, \phi) =&\sqrt{C_0} \hat{\bm \phi}\frac{1}{2\pi}\sum_n F^*(k_n) \notag \\
                            & \times  \int \limits_{-\pi}^{\pi}e^{i(m-n)\phi' -ik_0R\sin \theta \cos{(\phi'- \phi)}} \cos(\phi'-\phi)\, d\phi' \notag\\
                     =&\frac{\sqrt{C_0}\hat{\bm \phi}}{2}\sum_n F^*(k_{m+n})i(ie^{i\phi})^{-n} \notag  \\
                     & \times \Bigl( J_{n-1}(k_0R\sin \theta)  -J_{n+1}(k_0R \sin \theta) \Bigr ).
\end{align}
Similarly, we have
\begin{align}
\bm S_{-m}(\theta, \phi) =&\frac{\sqrt{C_0}\hat{\bm \phi}}{2}\sum_n F(k_{m+n})i(ie^{-i\phi})^{-n}  \\ \notag
                     & \times \Bigl( J_{n-1}(k_0R\sin \theta)  -J_{n+1}(k_0R \sin \theta) \Bigr ).
\end{align}
It follows from Eq.~(38) that
\begin{align}
& \Gamma_{m,m}= \frac{\pi C_0}{2} \sum_n \left| F(k_{m+n}) \right|^2 \notag \\
&               \quad  \times \int\limits_{0}^{\pi} \Bigl( J_{n-1}(k_0R\sin \theta)-J_{n+1}(k_0R \sin \theta) \Bigr )^2 \sin \theta \, d\theta,
\end{align}
and
\begin{align}
& \Gamma_{m,-m}=\frac{\pi C_0}{2} \sum_n  F(k_{m+n}) F(k_{m-n}) \notag \\
&                \quad \times \int\limits_{0}^{\pi} \Bigl( J_{n-1}(k_0R\sin \theta)-J_{n+1}(k_0R \sin \theta) \Bigr )^2 \sin \theta \, d\theta.
\end{align}

\section{Applications}
To demonstrate the applicability of the developed model, in this section we will apply it to three different examples. The first two examples deal with one and two scatterers, respectively, and in the third example we consider the fabrication-induced surface roughness present in high-$Q$ microdisk resonators, which corresponds to thousands of small scatterers. Numerical and experimental evidences are provided to support the derived theoretical results.
\subsection{Single scatterer}
For the single-scatterer case,  $\Delta\varepsilon_r(\phi)=\alpha \delta(\phi -\phi_0)$, where $\alpha$ is a constant and $\phi_0$ is the position of the scatterer. From Eq.~(59), we have $F(k_n)=\alpha \exp(-in\phi_0)$. It follows from Eqs.~(60) and (61) that $G_{m,m}=g_0 \alpha $, and $G_{m,-m}=g_0 \alpha \exp(-i2m\phi_0)$. For the TM-polarized WGM mode, $\Gamma_{m,m}$ is given by Eq.~(66) as
\begin{equation}
\begin{split}
\Gamma_{m,m}&=2\pi C_0 \alpha^2 \sum_n  \int\limits_{0}^{\pi}J^2_n(k_0R\sin \theta)\sin^3\theta \, d\theta,\\
            &= \frac{8}{3} \pi C_0 \alpha^2,
\end{split}
\end{equation}
where the following Bessel identity has been used (see Appendix A):
\begin{equation}
\sum_n J^2_n(x)=1.
\end{equation}
Similarly, $\Gamma_{m,-m}$ is obtained from Eq.~(67) as
\begin{equation}
\Gamma_{m,-m}=\frac{8}{3} \pi C_0 \alpha^2 e^{-i2m\phi_0}.
\end{equation}
$\eta$ defined in Eq.~(49) can then be calculated to be $\eta=\exp{(-i2m\phi_0)}$, which indicates that the CW and CCW modes are equally weighted in the eigenmodes $a_\pm$. Moreover, from Eqs.~(47) and (48), we have
\begin{align}
\omega_\pm=\omega_c - G_0 \pm G_0,\\
\gamma_\pm =  \Gamma_0 \mp \Gamma_0 \quad,
\end{align}
with
\begin{align}
G_0 \equiv g_0 \alpha; \quad \Gamma_0 \equiv \frac{8}{3} \pi C_0 \alpha^2.
\end{align}
Equations (75) and (76) reproduce the familiar results for the single-scatterer example, which are independent of the position of the scatterer (i.e., $\phi_0$). This stems from the fact that physically measurable scalar variables should be invariant with respect to the choice of the coordinate origin. Therefore, one can simply take $\phi_0 = 0 $ and arrive at the same results in Eqs.~(75) and (76).

The example mentioned in Fig.~1(b), which is equivalent  to a negative-dielectric-constant scatterer on the surface of a larger-radius microresonator, can also be analyzed. Assuming the scatterers have uniformly covered the surface of the microresonator except for a vacancy at $\phi_0 = 0 $,  we have $\Delta \varepsilon_r(\phi)= \sum_i \delta(\phi -\phi_i) -\delta(\phi)$ (for simplicity, we neglect a constant here, i.e., $\alpha =1$), where ${\phi_i }$ is a set of angles  representing the locations of these small scatterers (with the vacancy filled too, since we subtract it in the second term of $\Delta\varepsilon_r(\phi)$), which uniformly fall in the range of (0, 2$\pi$). The first term in $\Delta\varepsilon_r(\phi)$ only contributes to $F(k_0)$, and we have $F(k_n)= -1$   for $n \neq 0$. As a result, $G_{m,-m}= -G_0$ . Though $F(k_0)$ is a large number, it does not contribute to the scattering process, given that physically it corresponds to a uniform thin layer of dielectrics. Mathematically, from Eqs.~(66) and (67), we find that the weight coefficient for  $F(k_0)$ is proportional to an integral of $J_m(k_0R\sin \theta)$  . One important property of $J_n(x)$ is that its value is only significant when $|n| < |x|$. Because $m > k_0R$   ($m= k_0Rn_{\text{eff}}$, with $n_{\text{eff}}$  being the effective index of the WGM mode), the contribution of $F(k_0)$   to $\Gamma_{m, m}$ and $\Gamma_{m,-m}$   is negligible.  Hence,  $\Gamma_{m,m}$ and $\Gamma_{m,-m}$ are the same as those of the single-scatterer case given by Eqs.~(72) and (74), respectively. This leads to $\text{Re}\,(G^*_{m,-m}\Gamma_{m,-m})=-G_0 \Gamma_0 <0$, and $\gamma_+> \gamma_-$, as expected.

\subsection{Two scatterers}
We start with two identical scatterers. Since we have argued that only the relative positions of scatterers are important and the choice of the azimuthal origin can be arbitrary, we can take  $\Delta\varepsilon_r(\phi)= \delta(\phi - \phi_0)+ \delta(\phi +\phi_0)$ (again, we have omitted a constant in this expression). As a result, $F(k_n)=2\cos{n\phi_0}$, $G_{m,m}=2G_0$, and $G_{m,-m}=2G_0\cos{2m\phi_0}$. For the TM-polarized WGM mode,  it follows from Eq.~(66) that
\begin{align}
\Gamma_{m,m}= 8\pi C_0 \sum_n  & \cos^2 \left( (m+n)\phi_0 \right)  \notag \\
              & \times \quad \int\limits_{0}^{\pi}J^2_n(k_0R\sin \theta)\sin^3\theta \, d\theta.
\end{align}
The above result can be simplified using the following Bessel identities (see Appendix A):
\begin{align}
\sum_n J^2_n & (k_0R\sin \theta) \sin{2n\phi_0}=0 ,\\
\sum_n J^2_n(k_0R\sin \theta) & \cos{2n\phi_0}=J_0(2k_0R\sin \theta \sin \phi_0) ,
\end{align}
and we get
\begin{equation}
\Gamma_{m,m} =2\Gamma_0 \left ( 1+ p(2k_0R \sin \phi_0)\cos{2m\phi_0} \right),
\end{equation}
where $p(x)$ defined in Eq.~(35) is used. Similarly, from Eq.~(67),
\begin{equation}
\Gamma_{m,-m} =2\Gamma_0 \left ( p(2k_0R \sin \phi_0)+ \cos{2m\phi_0} \right).
\end{equation}
Equations (81) and (82) can also be derived from Eq.~(36), which is more convenient for this case. Since $G_{m,-m}$ and $\Gamma_{m,-m}$ are both real, from Eq.~(49), $\eta=\pm 1$.  By our convention, the choice of $\eta$ is to ensure $\omega_+$ is no less than $\omega_-$. Thus, according to Eq.~(47), $\eta$ takes $+1$ when $G_{m,-m} > 0$ and takes $-1$ when $G_{m,-m} < 0$, and can take either $1$ or $-1$ when $G_{m,-m} = 0$ , at which point the two eigenmodes share  the same resonance frequency. Equation (47) then becomes
\begin{equation}
\omega_\pm=\omega_c -2G_0 \pm 2G_0 |\cos{2m\phi_0}|.
\end{equation}
In addition, $\gamma_\pm$ is given by Eq.~(48) as
\begin{equation}
\begin{split}
\gamma_\pm = & \Gamma_{m,m} \mp \eta \Gamma_{m,-m}\\
           = & 2\Gamma_0 \left ( 1+ p(2k_0R \sin \phi_0)\cos{2m\phi_0} \right)\\
             & \mp 2\Gamma_0 \left ( |\cos{2m\phi_0}| + \text{sign}(\cos{2m\phi_0})p(2k_0R\sin \phi_0)\right),
\end{split}
\end{equation}
where in Eq.~(84) we have substituted $\eta$ by the sign function of $\cos{2m\phi_0}$, which only differs with $\eta$  at the zeros of $\cos{2m\phi_0}$. There are two reasons that allow us to do that. First, at the zeros of $\cos{2m\phi_0}$, the definition of $\gamma_\pm$ becomes ambiguous, since the subscripts $\pm$ are only used to distinguish  the resonance frequencies of the two eigenmodes. Second, as we shall show below, the zeros of $G_{m,-m}$ ($\varpropto \cos{2m\phi_0}$) are singular points of $\gamma_\pm$, where the behavior of $\gamma_\pm$ can only be studied by infinitely approaching these points.

As has been discussed in the comparison with the independent-scatterer approach, when the two scatterers are separated at large distances, $p(x)$ can be neglected in Eq.~(84) and we have
\begin{equation}
\gamma_\pm \approx 2\Gamma_0 \mp 2\Gamma_0 |\cos{2m\phi_0}|,
\end{equation}
which is identical with Eq.~(2) from the independent-scatterer model ($N = 2$). However, when the two scatterers are close to each other, $p(x)$ has to be considered in Eq.~(84). One special example is that the two scatterers overlap with each other (i.e., $\phi_0=0$\ and $p(0) = 1$), which can be treated as a single-scatterer case. Equation (84) then predicts $\gamma_+$ and $\gamma_-$ to be $0$ and $8\Gamma_0$, respectively. In contrast, Eq.~(2) provides $\gamma_+$ and $\gamma_-$ to be $0$ and $4\Gamma_0$, respectively. Using the result of Eq.~(76) for the single scatterer ($\gamma_+=0$\ and $\gamma_-=2\Gamma_0$), we find that Eq.~(84) is accurate and Eq.~(2) only predicts half of the exact number for $\gamma_-$(as seen from Eq.~(77), the scattering loss is proportional to the square of the dielectric perturbation so should be four times bigger if the dielectric perturbation doubles). Another interesting observation is that when approaching the zeros of the $\cos{2m\phi_0}$, $p(2k_0R\sin \phi_0)$ is generally nonzero, and
\begin{equation}
\gamma_\pm \approx 2\Gamma_0 \mp 2\Gamma_0 \text{sign}(\cos{2m\phi_0})p(2k_0R\sin \phi_0).
\end{equation}
If we sweep $\phi_0$ continuously, each time $\cos{2m\phi_0}$ crosses its zero points, its sign will change and there will be abrupt changes in $\gamma_+$ and $\gamma_-$, indicating the zeros of $G_{m,-m} (\propto \cos{2m\phi_0})$ are singular points of $\gamma_+$ and $\gamma_-$. Moreover, since the relation between $\gamma_+$ and $\gamma_-$ are reversed when passing the zeros of $\cos{2m\phi_0}$, it is always possible to observe $\gamma_+ > \gamma_-$ in the neighborhood of $G_{m,-m} = 0$ (as long as $p(x)$ is not negligible).

To verify the derived theoretical results, we perform a numerical investigation for a two-scatterer example using an in-house two-dimensional (2-D) microresonator mode solver implemented in the COMSOL  environment \cite{Comsol}. Details of the implementation are provided in Appendix B. The inset of Fig.~(4) illustrates the studied structure, which consists of two $10$-nm-radius scatterers attached to the surface of a $2$-$\mu$m-radius microdisk resonator. We fix the position of one scatterer and sweep the position of the other. The complex eigenfrequencies of the coupled system are computed by the mode solver, offering both the resonance frequencies and the scattering loss rates for the two eigenmodes. In Fig.~(4), two normalized parameters $\omega_{\text{diff}}$ and $\gamma_{\text{diff}}$ are plotted, which are defined as
\begin{align}
\omega_{\text{diff}} &\equiv \frac {\omega_+ -\omega_- }{4G_0},\\
\gamma_{\text{diff}} &\equiv \frac{ \gamma_+ -\gamma_-}{4\Gamma_0}.
\end{align}
\begin{figure}
\includegraphics {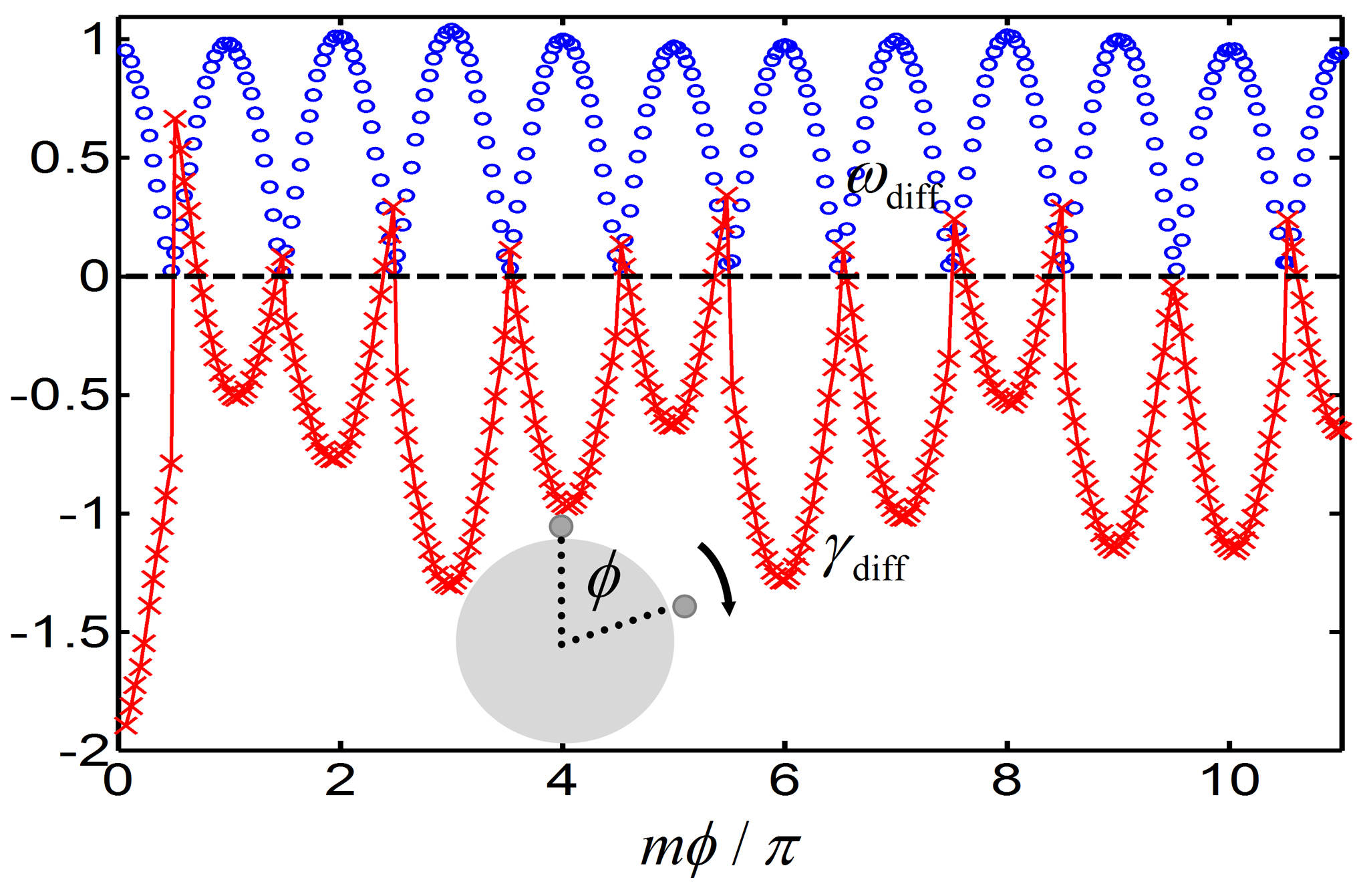}
\caption{ Simulation results of $\omega_{\text{diff}}$ and $\gamma_{\text{diff}}$, which are defined by Eqs.~(87) and (88), respectively, for two 10-nm-radius scatterers attached to the surface of a 2-$\mu$m-radius microdisk resonator as illustrated by the inset. The refractive index of the microdisk is 2.829 (obtained using the effective index method for a 220-nm-thick silicon layer), and the refractive index of the scatterers is twice as big  (i.e., $n_{\text{scatterer}}= 5.658$) to make the scattering effect significant. $m$ is the azimuthal order of the WGM mode (TM polarized), which is 19. $\omega_c$ is obtained from the simulation for an ideal microdisk resonator without any scatterers as 1.2282723e15 rad/s, and $G_0$ and $\Gamma_0$ are obtained from the single-scatterer simulation as 5.88e10 rad/s and 9.3e8 rad/s, respectively. }
\end{figure}

\noindent From Eqs.~(83) and (84), our theoretical model predicts
\begin{align}
\omega_{\text{diff}} = & \ |\cos{m \phi}|,\\
\gamma_{\text{diff}} = -|\cos{m\phi}|& - \text{sign}(\cos{m\phi}) p(\phi),
\end{align}
where the angular separation between the two scatterers are $\phi=2\phi_0$ as shown in the inset of Fig.~(4). Comparing Fig.~(4) to Eqs.~(89) and (90), we find $\omega_{\text{diff}}$ agrees with Eq.~(89) well; and $\gamma_{\text{diff}}$ indeed changes sign when passing the zeros of $\omega_{\text{diff}}$. We also notice that the magnitude of $\gamma_{\text{diff}}$  can be less than $-1$; especially, it approaches to $-2$ when the two scatterers are close to each other, as expected from the single-scatterer result. In Fig.~(5), we plot two additional normalized parameters $\omega_{\text{sum}}$ and $\gamma_{\text{sum}}$ defined as
\begin{align}
\omega_{\text{sum}} \equiv \frac{ \omega_+ + \omega_- -2\omega_c}{4G_0}, \\
\gamma_{\text{sum}} \equiv \frac{ \gamma_+ + \gamma_-}{4\Gamma_0}, \qquad
\end{align}
which are shown by the blue solid line and the red triangle marks, respectively. According to Eqs.~(83) and (84),
\begin{align}
\omega_{\text{sum}} & =-1, \\
\gamma_{\text{sum}} = 1+ & p(\phi)\cos{m \phi}.
\end{align}
\begin{figure}
\includegraphics {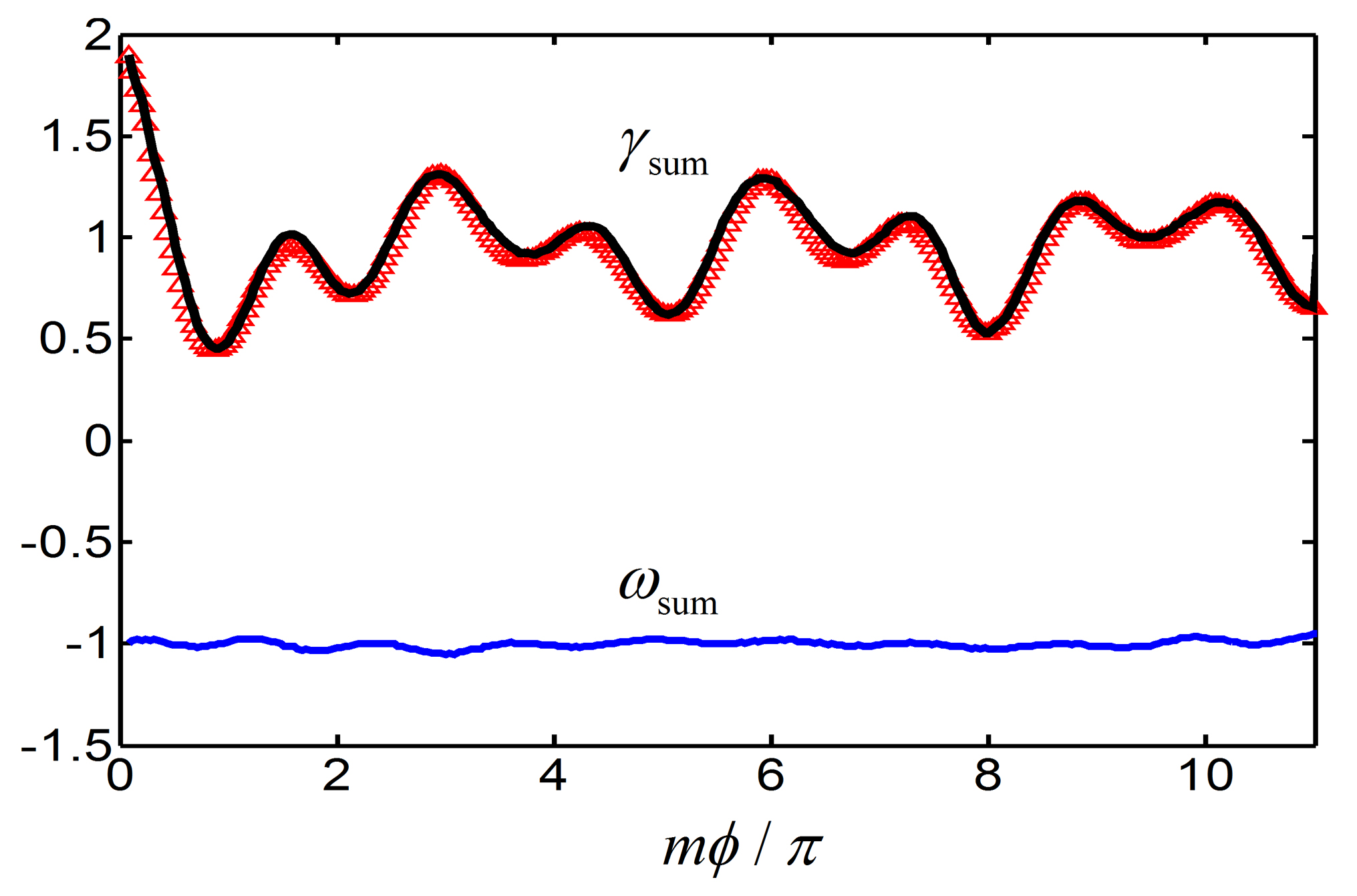}
\caption{ Numerical results of $\omega_{\text{sum}}$ and $\gamma_{\text{sum}}$, which are defined by Eqs.~(91) and (92), respectively. The red triangles corresponds to $\gamma_{\text{sum}}$ directly from simulation, while the black line corresponds to $\gamma_{\text{sum}}$ obtain by extracting $p(\phi)$ from the simulation result  of $\gamma_{\text{diff}}$ first (using Eq.~(90)) and then computing the numerical values of Eq.~(94).}
\end{figure}

\noindent As observed from Fig.~(5), $\omega_{\text{sum}}$ has a few percent fluctuations around the theoretical value (i.e., $-1$), largely arising from the limited positioning resolution of the moving scatterer when we sweep it along the perimeter of the microdisk ($1$ nm in the COMSOL environment). With the help of Eq.~(90), we could extract $p(\phi)$ from the numerical result of $\gamma_{\text{diff}}$ shown in Fig.~(4), and the result is depicted by the red triangles in Fig.~6. Moreover, using the obtained $p(\phi)$, $\gamma_{\text{sum}}$ could be computed based on Eq.~(94). The result, which is shown by the black solid line in Fig.~5, agrees with the one from direct simulation (red triangle marks) well, implying that our theoretical model is self-consistent. One may notice that $p(\phi)$ shown in Fig.~6 is different from the one plotted in Fig.~3(b). This is because $p(\phi)$ shown in Fig.~3(b) is for the three-dimensional (3-D) case, while our simulation considers a 2-D model. The essential difference can be traced back to the difference in the free-space Green's function \cite{Kong}. Employing the 2-D free-space Green's function and following a similar procedure as in the 3-D case(see the end of Appendix A), we obtain
\begin{equation}
p(\phi)=J_0(k_0d)=J_0\left( 2k_0R\sin{\frac{\phi}{2}}\right),
\end{equation}
\begin{figure}
\includegraphics {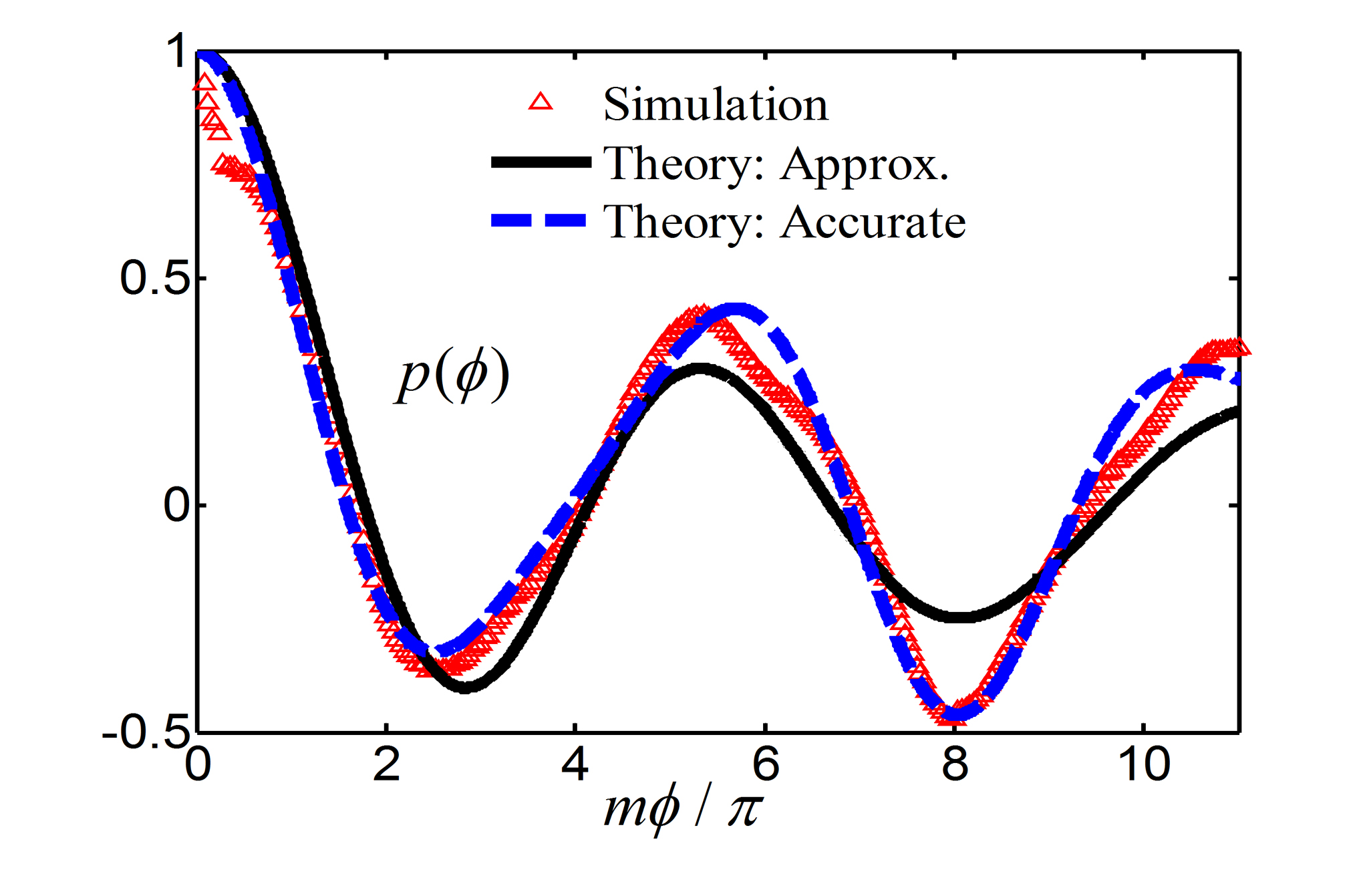}
\caption{ The red triangles corresponds to $p(\phi)$ extracted from the simulation result  of $\gamma_{\text{diff}}$ shown in Fig.~4 based on Eq.~(90). The black solid line is the numerical result of Eq.~(95), which is the theoretical prediction of $p(\phi)$ using the approximate 2-D Green's function. The blue dashed line is the numerical result of Eq.~(B11) (Appendix B), which is the theoretical prediction of $p(\phi)$ using the accurate 2-D Green's function.}
\end{figure}

\noindent where $d$ is the distance between the two scatterers. This result can also be expected from Eq.~(35), by taking the inclination coordinate $\theta = \pi /2$ and skipping the integration over $\theta$. In Fig.~6, we have plotted the predicted $p(\phi)$ given by Eq.~(95) by the black solid line, which agrees with the one extracted from the numerical simulation (shown by red triangles) reasonably well. The deviation there arises from two facts. First, we have certain positioning error when sweeping the scatterer in the simulation, as already mentioned for $\omega_{\text{sum}}$ in Fig.~5. Second, an accurate $p(\phi)$ requires taking the effect of the microdisk resonator to the free-space Green's functions into account, which has been omitted in Eq.~(95) (or Eq.~(35))(see the end of Appendix A for more discussions). In the Appendix B, a brief derivation is provided for the accurate calculation of $p(\phi)$ in the 2-D space, and the result is shown by the blue dashed line in Fig.~6, which agrees with the simulation result well. From the asymptotic behavior of $J_0(x)$ (Appendix A), one notice that in the 2-D case, the magnitude of $p(x)$ decreases with the separation distance $d$ of the two scatterers as $1/\sqrt{k_0d}$, instead of $1/k_0d$ as in the 3-D case (Fig.~3(b)). Therefore, a larger separation distance is required for 2-D models to neglect the effect of $p(\phi)$ ($d/\lambda_0> 8$ for $|p(\phi)| < 0.1$).

Finally, we would like to mention that if the two scatterers are not identical, by properly choosing the origin of the azimuth, we can still make $G_{m,-m}$ (which only depends on $F(k_{2m})$) real, but $\Gamma_{m,-m}$ (which depends on multiple terms of ${F(k_n)}$ as given by Eq.~(67)) generally will be complex. From Eq.~(49), $\eta$ does not necessarily have a magnitude of 1, which means the CW and CCW modes are not equally weighted in the eigenmodes. To observe a significant deviation of $|\eta|$ from 1, $|\Gamma_{m,-m}|$ has to be close to $|G_{m,-m}|$, which could only happen when $G_{m,-m}$ is near its zero points, since for dielectric scatterers $G_0 \gg \Gamma_0$ (as an example, for the scatterers studied in Fig.~4, $G_0 / \Gamma_0=63$). Thus, our model offers a simple explanation for the inequality of the CW and CCW components in the composition of the eigenmodes, which is studied in detail in Ref.~\cite{WGMstructure}.

\subsection{Fabrication-induced surface roughness}
Here we will examine a different example, i.e., the sidewall roughness caused by the imperfect fabrication of microresonators, which corresponds to numerous small scatterers on the surface. The distribution of these scatterers is random and usually follows a stationary statistic as \cite{Barwicz}
\begin{equation}
 <\Delta r(x) \Delta(x')>=\sigma^2 \exp{(- \frac{|x-x'|}{L_c} )},
\end{equation}
or
\begin{equation}
 <\Delta r(x) \Delta(x')>=\sigma^2 \exp{(- \frac{|x-x'|^2}{L^2_c} )},
\end{equation}
where $\Delta r(x)$  is the sidewall roughness along the wave propagation direction $x$;  $< >$ stands for the ensemble average; $\sigma$ is the roughness standard deviation; and $L_c$ is the correlation length. For the microdisk resonator, the dielectric perturbation function $\Delta\varepsilon_r(\phi)$ is related to the sidewall roughness as $\Delta\varepsilon_r(\phi)=\delta n^2 \Delta r(R\phi)$, where $\delta n^2=n^2_d -n^2_0$ , with $n_d$ and $n_0$ being the refractive indices of the microresonator and the surrounding medium (air in our case), respectively. Substituting Eq.~(58) into Eqs.~(96) and (97), we obtain
\begin{equation}
 <F(k_n)F^*(k_m)>=\frac{4\pi (\delta n^2)^2 \sigma^2L_c}{R\left( 1+ (k_nL_c)^2 \right)} \delta(n-m),
\end{equation}
and
\begin{equation}
\begin{split}
 <F(k_n)F^*(k_m)>=&\frac{2\pi^{\frac{3}{2}} (\delta n^2)^2 \sigma^2L_c}{R} \exp{(-(\frac{k_nL_c}{2})^2)} \\
                  & \times \delta(n-m),
\end{split}
\end{equation}
respectively. The Kronecker's delta function in both Eqs.~(98) and (99) implies that $[F(k_n)]$ ($n > 0$) are statistically independent random variables (remember $F(k_n)=F^*(k_{-n})$). Specifically, each resonator is one possible realization of $[F(k_n)]$, and Eqs.~(98) and (99) are valid when the ensemble average is performed for many independently fabricated resonators under the same condition.

The independence of $[F(k_n)]$ ($n > 0$) provides key insights to the understanding of the properties of microdisk resonators such as the one shown in Fig.~2. Using the results of Eqs.~(52)-(54) under the assumption of $|G_{m,-m}| \gg |\Gamma_{m,-m}|/2$, we find the mode splitting for the azimuthal order $m$ is proportional to $|G_{m,-m}|$ and therefore $|F(k_{2m})|$ (Eqs.~(53) and (61)). Because of the independence of $[F(k_n)]$, different azimuthal orders can have independent mode splittings. Strong variations of mode splitting over azimuthal orders for an individual microresonator are thus possible if the corresponding $[F(k_n)]$ have strong amplitude variations with the index $n$. Moreover, $[F(k_n)]$ should also have independent phase variations. For example, if $[F(k_n)]$ are all positive numbers, then the correlation function in Eqs.~(98) and (99) will also be positive, contradicting with the results there. The variations of the phase and amplitude of $[F(k_n)]$ (with the index $n$) can lead to variations of scattering losses for different azimuthal orders. To see this, we use the TM polarization as an example. In Eqs.~(66) and (67), $\Gamma_{m,m}$ and $\Gamma_{m,-m}$ for the azimuthal order $m$ is summed over terms of $[F(k_n)]$ with indices around $m$, weighted by coefficients as integrals of $J_n(k_0R\sin \theta)$. As already mentioned, $J_n(x)$ only has significant values when $|n| < |x|$. Therefore, for azimuthal order $m$, the summation in Eqs.~(66) and (67) for $\Gamma_{m,m}$ and $\Gamma_{m,-m}$ only contain limited terms  of $[F(k_n)]$, with contributing indices in the range of $(m-k_0R, m+k_0R)$ (see Fig.~7 for an illustration). This has two consequences. First, $\Gamma_{m,m}$ and $\Gamma_{m,-m}$ will show some dependencies on $m$ because of this local summation cannot average out the variations among $[F(k_n)]$. Second, whether $\gamma_+ > \gamma_-$ or $\gamma_+ < \gamma_-$ is determined by $\text{Re}\,(G^*_{m,-m}\Gamma_{m,-m})$, which is given by
\begin{equation}\
\qquad \text{Re}\,(G^*_{m,-m}\Gamma_{m,-m}) \varpropto \text{Re}\,(F^*(k_{2m})\Gamma_{m,-m}).
\end{equation}
Since $m > k_0R$, $F(k_{2m})$ does not correlate with $\Gamma_{m,-m}$ (see Fig.~7). If the phase of $F(k_{2m})$ can vary randomly within $(0, 2\pi)$, then $\text{Re}\,(G^*_{m,-m}\Gamma_{m,-m})$ can be either positive or negative. Furthermore, because of the independence $[F(k_n)]$, $\text{Re}\,(G^*_{m,-m}\Gamma_{m,-m})$ will also be independent for different azimuthal orders. For that reason, one can observe simultaneous occurrence of different lineshapes among the same radial mode family in an individual microresonator, as shown in Fig.~2. The above discussions also apply to the TE polarization, since the weight coefficients in Eqs.~(70) and (71) have similar properties as those of Eqs.~(66) and (67).
\begin{figure}
\includegraphics {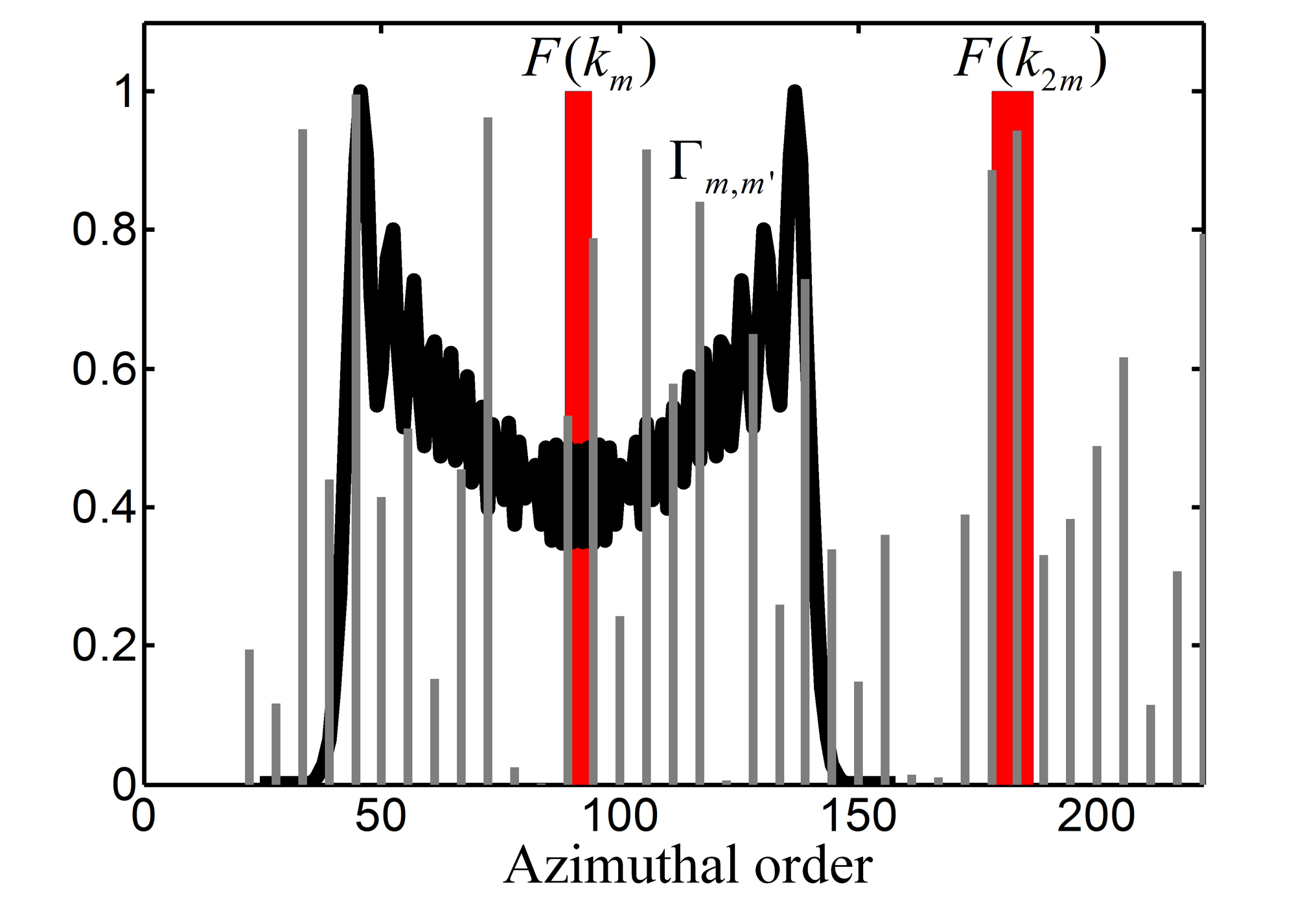}
\caption{ The grey thin bars depict $[F(k_n)]$, and the two red thick bars correspond to $F(k_m)$ and $F(k_{2m})$ for the azimuthal order $m$ under consideration. $\Gamma_{m,m}$ and $\Gamma_{m,-m}$, given by Eqs.~(66) and (67), respectively, are limited sums of  $[F(k_n)]$ with indices around $m$, with the weight coefficients given by the black solid line. On the other hand, $G_{m,-m}$ is proportional to $F(k_{2m})$, and therefore has no overlap with  $\Gamma_{m,m}$ or  $\Gamma_{m,-m}$.}
\end{figure}

The next question is what the amplitude and phase variations of $[F(k_n)]$ with the index $n$ for an individual microresonator. One method is to generate scatterer distributions following the statistical rule given by Eq.~(96) or (97), and calculate $[F(k_n)]$ according to Eq.~(59). We have constructed one such scatterer distribution in Appendix C, which satisfies Eq.~(96). Numerical simulations there indicate that the amplitude of each $F(k_n)$ follows a Rayleigh distribution and its phase follows a uniform distribution within $(0,2\pi)$. A more convenient way is to assume some simple statistical models for $[F(k_n)]$. For example, for the surface roughness that can be approximated by Eq.~(97), in Ref.~\cite{qingol} we have assumed a Gaussian distribution for  the amplitude and uniform distribution for the phase of $F(k_n)$ as
\begin{equation}
\begin{split}
 &F(k_n)  = \sqrt{\frac{2\pi^{\frac{3}{2}} (\delta n^2)^2 \sigma^2L_c}{R} \exp{\left(-(\frac{k_nL_c}{2})^2 \right)}} \\
                 \quad  &\times (\cos \alpha + \sin \alpha \, N_n(0,1)) \exp{(i 2\pi  U_n(0,1))},
\end{split}
\end{equation}
where $[N_n(0,1)]$ are independent random variables with a normal distribution with a zero mean and a unit variance, and $[U_n(0,1)]$ are independent random variables with a uniform distribution in the interval $(0, 1)$. The parameter $\alpha$ is introduced to account for the amplitude variations of $[F(k_n)]$. The independent uniformly distributed phase terms $\left[\exp{(i2\pi U_n(0,1))}\right]$ will ensure the independence of $[F(k_n)]$, as required by Eq.~(99).
With the generated $[F(k_n)]$, $G_{m,m}$, $G_{m,-m}$, $\Gamma_{m,m}$ , and $\Gamma_{m,-m}$ are obtained for each azimuthal order $m$, and the mode splitting and scattering loss rates are calculated from Eqs.~(47)-(49).
\begin{figure}
\includegraphics {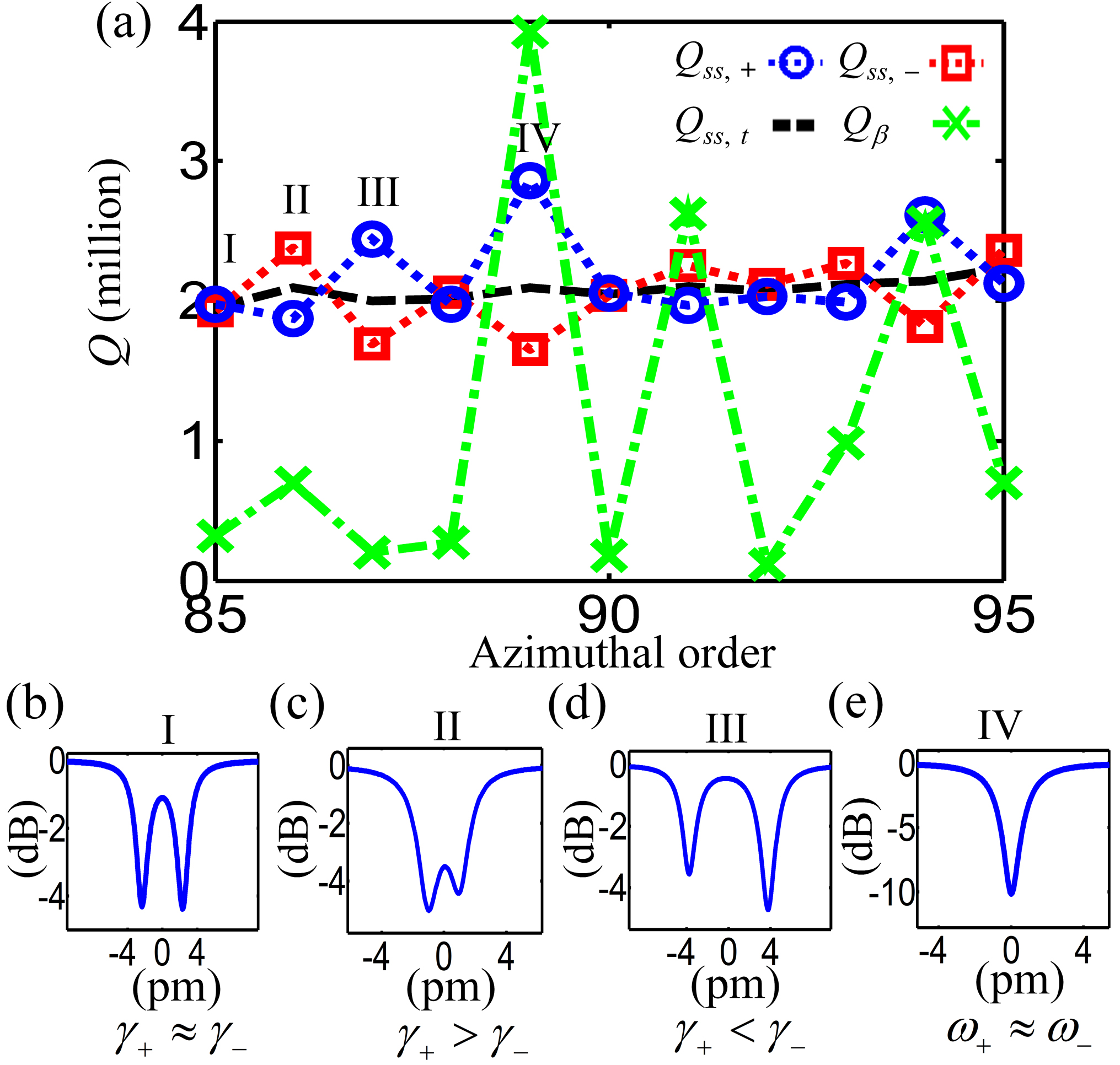}
\caption{(a) Simulated mode splitting and scattering loss for a 10-$\mu$m-radius silicon microdisk resonator with $[F(k_n)]$ generated from Eq.~(101). The parameters used are $\sigma = 2$ nm, $L_c=160$ nm, and $\alpha= 0.5 \pi$. $\left|\bar{\bm E}_m(R,0)\right|$ is evaluated by obtaining the fundamental TE WGM mode from 3-D finite element method (FEM) simulations \cite{mohammadoe}, followed by a subsequent averaging along the vertical dimension of the microdisk slab. The dependence of $\left|\bar{\bm E}_m(R,0)\right|$ on $m$ (or the wavelength) is small and can be neglected. (b)-(e) Transmission responses of the marked resonant modes with a coupling $Q$ of 600,000 (over-coupled). Therefore, similar to Fig.~2, a higher extinction on resonance indicates a broader linewidth.}
\end{figure}

Figure 8 shows one simulation example for a 10-$\mu$m-radius silicon microdisk resonator. We have defined dimensionless quality factors for the scattering loss as $Q_{ss,\pm} \equiv \omega_c/ \gamma_\pm$ and for the mode splitting as $Q_\beta \equiv \omega_c/ (\omega_+ - \omega_-)$. We also define a scattering $Q$ for the CW (CCW) mode as $Q_{ss,t} \equiv \omega_c/ \Gamma_{m,m}$, though it is not directly measurable since the CW and CCW modes are no longer the eigenmodes of the system. However, from Eq.~(48), we have
\begin{equation}
 2Q^{-1}_{ss,t}=Q^{-1}_{ss,+}+Q^{-1}_{ss,-}.
\end{equation}
In Fig.~8(a), we have plotted $Q_{ss,\pm}$ for the two eigenmodes in the blue dotted line with circles and red dotted line with squares, respectively;  $Q_{ss,t}$ is plotted in the black dashed line while $Q_\beta$ is shown by the green dash-dotted line with crosses. We have chosen the parameters in Eq.~(101) to generate close results to the experimental ones as shown in Fig.~2, with $\sigma=2$ nm, $L_c =160 $ nm, and $\alpha=0.5 \pi$. As observed from Fig.~8(a), indeed we have strong variations of mode splitting for different azimuthal orders. Furthermore, as shown by the four zoom-in figures in Figs.~8(b)-(e), different lineshapes can be observed for different azimuthal orders in the transmission measurement, similar to the experimental results shown in Figs.~2(b)-(e). In particular, in Fig.~8(e), only a single resonance is observed because the mode splitting is negligible. However, the resonance actually corresponds to two eigenmodes with different scattering loss rates, and will show different responses under different coupling conditions. Another point worth mentioning is that the scattering $Q$s of the both eigenmodes exhibit more than $30\%$ variations over azimuthal orders, and in a scattering-loss limited microdisk resonator, such variations will be transferred to the intrinsic $Q$s \cite{borselliol,qingol,mohammadoe} . Finally, because of the random nature of Eq.~(101), Fig.~8 is just one possible result, and different simulation runs will generate close but not exactly the same outcomes. This in fact closely mimics the real fabrication, which produces resonators with comparable but rarely identical performances.
\section{Conclusions}
In summary, we have developed a unified model that applies to an arbitrary number of scatterers, which provides a comprehensive understanding on the mode splitting and scattering loss in high-$Q$ WGM microresonators. Compared with the independent-scatterer approach which is commonly used for the a-few-scatterer scenario, our work reveals that the independent-scatterer model has neglected the interference terms from different scatterers, whose effect decreases with the separation distance $d$ as $1/k_0d$ for 3-D cases and as $1/\sqrt{k_0d}$ for 2-D cases. Thus, the independent-scatterer model only works when scatterers are well separated. Compared with the intuitive physics approach which is developed for the many-scatterer scenario, we have derived an additional coupling term between the CW and CCW modes (i.e.,\ $\Gamma_{m,-m}$) that has been missing in the phenomenological model used by the intuitive physics approach. This modification leads to the prediction of asymmetric lineshapes in a self-consistent manner. Moreover, combined with numerical studies and experimental results, the unified model has provided many new understandings on the mode splitting and scattering loss in high-$Q$ WGM microresonators. For example, we prove that the intuitive belief that $\gamma_+ \leq \gamma_-$ is not generally true, and counter examples can even be found for two scatterers attached to the surface of WGM microresonators. Our work also unveils  that when mode splitting disappears, the scattering loss rates of the two eigenmodes are generally different, and $\gamma_+$ and $\gamma_-$ become singular at these points. For the fabrication-induced surface roughness present in high-$Q$ microresonators, the stationary distribution of numerous small scatterers results in independent mode splitting for different azimuthal orders. The scattering loss rate of each eigenmode will also show more than 30\% variations among the radial mode family, which explains the observed variations of intrinsic $Q$s in scattering-loss-limited microresonators \cite{qingol,borselliol}. We believe such a unified approach does not only fill the gap for the existing theoretical works on the mode splitting and scattering loss in high-$Q$ WGM microresonators, but also will play an indispensable role for the practical applications of these phenomenons to produce the most accurate results.

\begin{acknowledgments}
We would like to thank helpful discussions with Dr. Mohammad Soltani. This work is supported by the DARPA Microsystems Technology Office (MTO) under grant No.~2106ATG.
\end{acknowledgments}

\appendix

\section{Mathematical Formulas}
In this part, we list the mathematical formulas that are used in the paper, mainly for the Bessel functions. Brief derivations are provided for some equations which are not easily found in mathematical handbooks.

$J_n(x)$, the Bessel function of the first kind, has the following integral representation \cite{math}:
\begin{equation}
J_n(x)=\frac{1}{2\pi} \int\limits_{-\pi}^{\pi}e^{i(x\sin\tau -n\tau)}\,d\tau.
\end{equation}
Substituting $\tau=\pi/2 -\tau'$, we obtain another representation of $J_n(x)$ as
\begin{equation}
J_n(x)=\frac{1}{2\pi} i^{-n}\int\limits_{-\pi}^{\pi}e^{i(x\cos\tau \pm n\tau)}\,d\tau.
\end{equation}
When $x \gg n^2$, we have the following asymptotic forms \cite{math}:
\begin{align}
J_n(x) & \approx \sqrt{\frac{2}{\pi x}} \cos{(x - \frac{n \pi}{2} - \frac{\pi}{4})}, \\
H^{(1)}_n(x) & \approx \sqrt{\frac{2}{\pi x}} \exp{(i(x - \frac{n \pi}{2} - \frac{\pi}{4}))},
\end{align}
where $ H^{(1)}_n(x)$  is the Hankel function of the first kind.

To obtain $p(x)$ defined in Eq.~(35), we employ Sonine's first integral as \cite{Bessel}
\begin{equation}
\int\limits_{0}^{\pi/2} J_0(x\sin \theta)\sin \theta \cos^{2\nu +1}\theta \, d\theta = \frac{2^\nu \Gamma{(\nu +1)}J_{\nu+1}(x)}{x^{\nu +1}},
\end{equation}
where $\Gamma(\nu+1)$ is the gamma function \cite{math}. $p(x)$ can be evaluated by taking $\nu$ in Eq.~(A5) to be $-\frac{1}{2}$ and $\frac{1}{2}$ and subtracting these two terms, which results in
\begin{equation}
 p(x)=\frac{3}{2}\Bigl  (\sqrt{\frac{\pi}{2x}}J_{1/2}(x)-\frac{1}{x}\sqrt{\frac{\pi}{2x}}J_{3/2}(x) \Bigr).
\end{equation}
Equation (A6) can be further simplified with the help of the spherical Bessel functions \cite{math}, and we obtain
\begin{equation}
p(x)=\frac{3}{2}\Bigl( \frac{\sin x}{x} + \frac{\cos x}{x^2} - \frac{\sin x}{x^3}   \Bigr).
\end{equation}
When $x$ is large, the leading term of $p(x)$ is the sinc function and its envelop varies as $3/2x$.

Equations (73), (79), and (80) can be proved by considering the following series, which can be calculated using the integral representation of $J_n(x)$ given by Eq.~(A1) as
\begin{equation}
\begin{split}
\sum_n & e^{i2n \phi_0} J^2_n(x) =  \frac{1}{4\pi^2}\iint  \\
                   & \sum_ne^{i(x \sin \tau_1 + x\sin \tau_2 -n\tau_1 -n\tau_2 +2n\phi_0)} \, d\tau_1 d\tau_2.
\end{split}
\end{equation}
Using the following identity:
\begin{equation}
 \delta(\phi)=\frac{1}{2\pi}\sum_n e^{in\phi},
\end{equation}
we have
\begin{equation}
\begin{split}
\sum_n & e^{i2n \phi_0} J^2_n(x) = \frac{1}{2\pi} \\
                   & \iint e^{i(x \sin \tau_1 + x\sin \tau_2) } \delta ( \tau_1 + \tau_2 -2 \phi_0) \, d\tau_1 d\tau_2.
\end{split}
\end{equation}
Changing the integration variables from $\tau_1$  and $\tau_2$ to $(\tau_1 + \tau_2)/2$    and $(\tau_1 - \tau_2)/2$  would lead us to
\begin{equation}
 \sum_n e^{i2n \phi_0} J^2_n(x) = J_0(2x \sin{\phi_0}).
\end{equation}
In particular, if we let $\phi_0 = 0$, Eq.~(A11) becomes
\begin{equation}
\sum_n J^2_n(x)=1.
\end{equation}

Regarding the calculation of $p(x)$ for the 2-D space, we start with the 2-D free-space Green's function $H^{(1)}_0(k_0r)$ \cite{Kong}. For the TM polarization,  the far-field electric field is given by
\begin{equation}
 \bm{E}_m^{\text{far}}(\bm{r}) \propto  \int \Delta\varepsilon(\bm{r}')\bm{E}_m(\bm{r}')H^{(1)}_0(k_0 |\bm r -\bm{r'}|)\, d^2 \bm r'.
\end{equation}
Using the asymptotic form of the Hankel function given by Eq.~(A4), we have
 \begin{equation}
  \bm{E}_m^{\text{far}}(\bm{r}) \propto  \frac{1}{\sqrt{r} }\int \Delta\varepsilon(\bm{r}')\bm{E}_m(\bm{r}') e^{-ik_0 \hat{\bm r} \cdot \bm r'}\, d^2 \bm r'.
 \end{equation}
 Comparing Eq.~(A14) with Eq.~(8), one can expect that for the 2-D case, the geometric integral in Eq.~(33) would be (for the TM polarization)
 \begin{equation}
  \int\limits_{-\pi}^{\pi} e^{ik_0 \hat{\bm k} \cdot (\bm x_n -\bm x_{n'}) } \, d\phi,
 \end{equation}
 which is just $J_0(k_0|\bm x_n -\bm x_{n'}|)$.

 A final note on the transition from Eq.~(19) to Eq.~(21). The simplification takes advantage of the following property of the spatial part of the electric field:
\begin{equation}
\nabla \times (\nabla \times \bm E(\bm r)) -\omega_c^2 \mu \varepsilon (\bm r) \bm E(\bm r) =0.
\end{equation}
Obviously, $\bm E_m(\bm r)$ satisfies the above equation, but $[\bm E_j(\bm r)]$ do not. In this paper, for $[\bm E_j(\bm r)]$, we have approximated $\varepsilon (\bm r)$ as $\varepsilon_0$, i.e., the effect of the microresonator to the Green's functions has been neglected, which is the essence of the volume current method that results in much simplified solutions with acceptable accuracies (a more physical explanation is that a proper choice of radiation modes should ensure that they are orthogonal to the WGM modes so they do not couple to the WGM modes without scatterers.  However, $[\bm E_j(\bm r)]$ given by Eq.~(18) do not satisfy this property). Therefore, a more rigorous calculation should take the effect of the microresonator into account, and in Eq.~(33) the exact Green's functions have to be used. For 2-D cases, such a task is relatively easy, and one illustrative example is provided in Appendix B. However, for 3-D cases, the accurate calculation of the Green's functions is usually difficult \cite{Kong}. 

\section{2-D FEM Simulation and Discussions}
In this part, we will describe the simulation method to obtain the complex eigenfrequencies of a 2-D microdisk resonator with scatterers attached on its surface. We use the finite element method (FEM), which is generally much faster compared with the finite-difference time-domain (FDTD) method. To obtain the scattering loss, perfectly matched layers (PMLs) are implemented based on the stretched coordinate method \cite{FDTD}. For example, in the cylindrical coordinate system, for the TM polarization, we have \cite{PML}
\begin{equation}
 \left[ \frac{\partial}{\rho \,\partial \rho}(\rho \frac{\partial}{\partial \rho} ) + \frac{1}{\rho^2} \frac{\partial^2}{\partial \phi^2}\right ] E_z=-k_0^2 n^2s^2_\rho E_z,
\end{equation}
where $n$ is the refractive index at each region; $s_\rho$ is the complex coordinate stretching factor for the PML (light is only attenuated in the increasing $\rho$ direction); and $k_0$ is the eigenvalue we try to obtain, which is related to the complex eigenfrequency $\omega$ as $\omega=k_0c$.

Equation (B1) is implemented and solved as a partial differential equation (PDE) in the commercial software COMSOL \cite{Comsol} . Because COMSOL does not provide the cylindrical coordinate system for structures without axial symmetry, Eq.~(B1) is converted back to the Cartesian coordinate system as
\begin{equation}
\left [\frac{\partial^2}{\partial x^2} +\frac{\partial^2}{\partial y^2} \right]E_z=-k_0^2n^2s^2_{\rho}E_z.
\end{equation}
Figure 9 shows the structure we simulate, where the microdisk is centered at the origin. $s_{\rho}$ is chosen to be the following form for the PML region:
 \begin{equation}
s_{\rho}=1+ ia \frac{(\sqrt{x^2 +y^2 } -\rho_0)^2}{d^2},
 \end{equation}
 where $\rho_0$ and $d$ are the starting radius and the thickness of the PML region, respectively, and $a$ is a parameter that can be adjusted for the optimum performance of PML (we take $a = 3$ in our simulation). $s_{\rho}$ is 1 for other regions.

 Similarly, for the TE polarization, the equation can be implemented as
 \begin{equation}
 \left [\frac{\partial}{\partial x}
 \left (\frac{\partial}{n^2 \partial x} \right )+ \frac{\partial}{\partial y}
  \left (\frac{\partial}{n^2 \partial y} \right ) \right ]H_z=-k_0^2s^2_{\rho}H_z.
 \end{equation}
 The placement of $n^2$ inside the first-order derivative is to ensure correct boundary conditions when the PDE is solved (i.e., $E_\phi$ to be continuous) \cite{PML2}.

 \begin{figure}
 \includegraphics[scale=0.8]{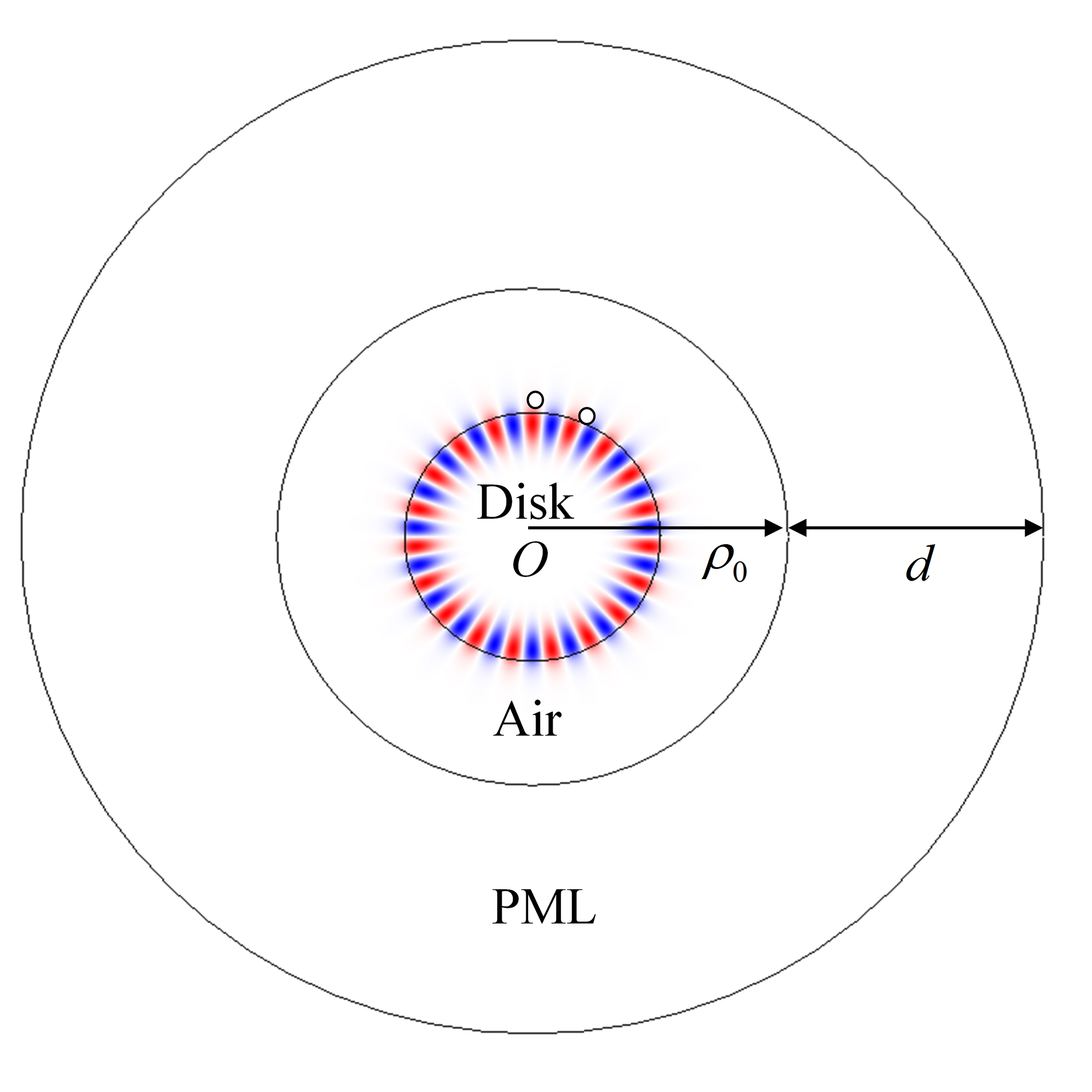}
 \caption{ Simulated structure in COMSOL. The two small scatterers have been exaggerated in size for the illustration purpose.}
 \end{figure}

 For FEM simulations, mesh size has to be small enough to avoid artificial effects. In our case, cubic meshes with grid size less than 50 nm are employed (of course, for areas surrounding the scatterers the mesh has to be finer). The memory requirement is not that demanding (desktop computers with 8 Gb memory run the simulation smoothly), and each simulation typically takes a few minutes. Moreover, the scan of the position of the scatterer can be facilitated with the use of the COMSOL Script (or COMSOL with MATLAB)\cite{Comsol}.

 As a supplementary discussion, here we provide a brief derivation for computing the accurate $p(\phi)$ in the 2-D space. Instead of using the approximate Green's function $H^{(1)}_0(k_0r)$ which neglects the effect of the microresonator (see discussions at the end of Appendix A),we seek the exact Green's function by solving the following equation (for TM-polarization)\cite{Kong}: 
 \begin{equation}
  \left[ \frac{\partial}{\rho \,\partial \rho}(\rho \frac{\partial}{\partial \rho} ) + \frac{1}{\rho^2} \frac{\partial^2}{\partial \phi^2} + k_0^2 n^2 \right] g(r,\phi)= \delta(r-R)\delta(\phi).
 \end{equation}
The solution has the following form \cite{borselli}: 
\begin{align}
g(r,\phi)&=\sum_n a_n J_n(k_0n_dr)e^{in\phi}\ \ \ \ \text{for} \ r<R,  \\
g(r,\phi)&=\sum_n b_n H^{(1)}_n(k_0n_0r)e^{in\phi} \ \ \text{for} \ r>R,
\end{align}
where $n_d$ and $n_0$ are the refractive indices of the microdisk and the surrounding medium (air here), respectively, and $a_n$ and $b_n$ are the corresponding expansion coefficients. Applying the boundary conditions of $g(r,\phi)$ at $r=R$ gives
\begin{align}
a_n J_n(k_0n_dR)  =b_n H^{(1)}_n(k_0n_0R), \qquad \quad \\
k_0 n_d a_n J'_n(k_0n_dR) - k_0 n_0 b_n H'^{(1)}_n(k_0n_0R) = \frac{1}{2\pi},
\end{align}
which yield
\begin{widetext}
\begin{equation}
b_n= \frac{ J_n(k_0n_dR)}{2\pi k_0 \left( n_d H^{(1)}_n(k_0n_0R) J'_n(k_0n_dR)- n_0  H'^{(1)}_n(k_0n_0R) J_n(k_0n_dR) \right) }.
\end{equation}
\end{widetext}
Inserting the obtained Green's function into Eq.~(33), one can easily derive $p(\phi)$ for the TM-polarized light as
\begin{equation}
p(\phi)=\frac{\sum_n |b_n|^2 \cos{n\phi}}{\sum_n|b_n|^2},
\end{equation}
where $\phi$ is the angular separation between the two scatterers as defined in the inset of Fig.~4. The numerical result of Eq.~(B11) for the example studied in Fig.~4 is shown by the blue dashed line in Fig.~6.

\section{Scatterer Distribution}
 In this part, we will construct a distribution of small scatterers that satisfies the statistical rule given by Eq.~(96). We consider identical scatterers, with the shape shown in Fig.~10(a). The height $\sigma$ and the width $W$ of each scatterer are assumed to be much smaller than the wavelength. We divide the perimeter of the microdisk resonator by $N$ divisions to allow for $N$ scatterers ($W= 2\pi R/N$), and each scatterer can be either pointing outward or inward, with a parameter $x_k$ being $+1$ for the former and $-1$ for the latter for the $k$th scatterer. A set of $[x_k]$ ($k = 1, 2, \cdots, N$) then describes the scatterer distribution on the perimeter of the microdisk resonator , and
 \begin{equation}
  \Delta r(\phi)=\sigma x_k, \quad \text{with}\ k=\text{Round}\, (\frac{\phi N}{2 \pi} ),
 \end{equation}
 where Round() denotes the nearest integer function \cite{math}. We generate $[x_k]$ using the following statistical rule:
\begin{equation}
x_k \in \{-1, 1\}, \ P(x_{k+1}=x_k)=\frac{1}{2}(1+\chi),
\end{equation}
where $\chi$ falls in the range between (0, 1). It is easy to verify
\begin{equation}
 E(x_kx_{k+n})=\chi^{|n|}, \quad \forall k,
\end{equation}
which states that the correlation between $x_k$ and $x_{k+n}$ only depends on their index difference. From Eq.~(C1),
\begin{equation}
<\Delta r(\phi) \Delta r(\phi + \phi')>=\sigma^2 E(x_kx_{k+n}) \approx \sigma^2 \chi^{\frac{|\phi'|N}{2\pi}},
\end{equation}
where we have approximated the index difference $n$  corresponding to $\phi'$ phase shift as $n \approx \phi'N/2\pi$, which is valid when $|n|$ is much larger than 1. As a result,
\begin{equation}
<\Delta r(\phi) \Delta r(\phi + \phi')> \approx \sigma^2 \exp{\left(-\ln {({\chi}^{-1})} \frac{|\phi'|N}{2\pi} \right)}.
\end{equation}
Comparing Eq.~(C5) to Eq.~(96), we find they have similar expressions (remember $x=R\phi$), and the correlation length  $L_c$ can be identified as
\begin{equation}
L_c=\frac{2\pi R}{N \ln{(\chi^{-1})}} = \frac{W}{\ln{(\chi^{-1})}}.
\end{equation}
$[F(k_n)]$ can then be calculated from Eq.~(59) as
\begin{equation}
F(k_n) \approx \frac{2\pi \delta n^2 \sigma}{N} \sum_k x_k e^{-i2\pi kn/N} =\frac{W \delta n^2 \sigma}{R} X_n,
\end{equation}
where $[X_n]$ are the DFT (discrete Fourier transform) of $[x_k]$.

\begin{figure}
\includegraphics {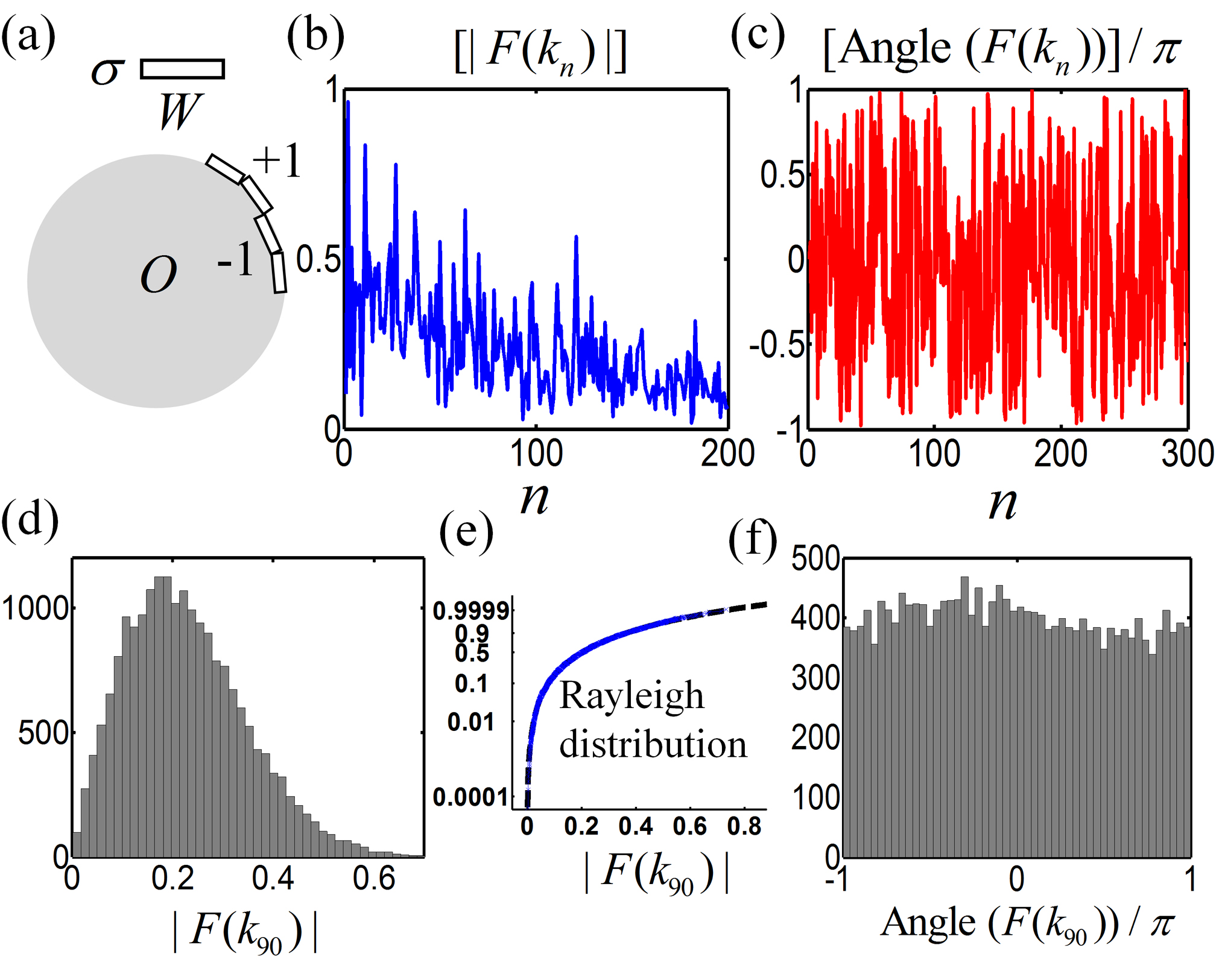}
\caption{(a) The scatterers considered in this model are all identical, and can point either outward ($+1$) or inward ($-1$) on the surface of the microresonator.(b)-(c) Amplitude and phase values of one simulation result of $[F(k_n)]$ ($R$= 10 $\mu$m, $N=5000$,\,$\chi=0.926$ so $L_c =160$ nm. For simplicity, we have assumed $\delta n^2 \sigma =1 $ in Eq.~(C7)). Strong variations of $[F(k_n)]$ with the index $n$ can be observed. (d)-(f) The value for a specific $F(k_n)$ ($n=90$) is recorded for multiple generated $[x_k]$ (20,000 runs). The histograms show that the amplitude of $[F(k_n)]$ follows a Rayleigh distribution (Fig.~10(d)) and the phase of $[F(k_n)]$ follows a uniform distribution in ($-\pi$, $\pi$) (Fig.~10(f)). Figure 10(e) plots of the cumulative distribution function of the obtained $|F(k_n)|$ (blue solid line) versus a fit for the Rayleigh distribution (black dashed line), where the $y$ axis is shown using the logarithmic scale.}
\end{figure}

Numerical experiments are performed in MATLAB by generating $[x_k]$ based on Eq.~(C2) and computing $[F(k_n)]$ from Eq.~(C7). Figures 10(b)-(c) show one example of $[F(k_n)]$, which confirm that there are strong amplitude (Fig.~10(b)) and phase (Fig.~10(c)) variations with the index $n$. In Figs.~10(d)-(f), we monitor the value of one $F(k_n)$ ($n = 90$) for each generated $ [{x_k}] $, and the histograms imply that the amplitude distribution of $F(k_n)$ is a Rayleigh distribution while the phase distribution of $F(k_n)$ is a uniform distribution in $(-\pi,\pi)$.


\bibliography{mode}

\end{document}